\documentclass[twocolumn]{aastex62}
\usepackage{amsmath}
\usepackage{natbib}
\usepackage{graphicx}

\usepackage{verbatim}

\shorttitle{Radiative Transfer Disk Masses}
\shortauthors{Ballering \& Eisner}

\newcommand{\NumFinalTargets}{132}

\newcommand{\PercentMdRadTranAboveMMSNnoconstraint}{31}
\newcommand{\PercentMdAnaAboveMMSNyesconstraint}{18}
\newcommand{\PercentMdRadTranAboveMMSNyesconstraint}{28}

\newcommand{\NumExcludedTargets}{five }
\newcommand{\ExcludedTargetsList}{DH Tau A, J04213459+2701388, J04330945+2246487, UX Tau A, and V892 Tau AB}

\begin{document}

\title{Protoplanetary Disk Masses from Radiative Transfer Modeling: A Case Study in Taurus}

\author{Nicholas P. Ballering}
\affiliation{Steward Observatory, University of Arizona, 933 North Cherry Avenue, Tucson, AZ 85721, USA}

\author{Josh A. Eisner}
\affiliation{Steward Observatory, University of Arizona, 933 North Cherry Avenue, Tucson, AZ 85721, USA}

\correspondingauthor{Nicholas P. Ballering}
\email{ballerin@email.arizona.edu}

\begin{abstract}
Measuring the masses of protoplanetary disks is crucial for understanding their planet-forming potential. Typically, dust masses are derived from (sub-)millimeter flux density measurements plus assumptions for the opacity, temperature, and optical depth of the dust. Here we use radiative transfer models to quantify the validity of these assumptions with the aim of improving the accuracy of disk dust mass measurements. We first carry out a controlled exploration of disk parameter space. We find that the disk temperature is a strong function of disk size, while the optical depth depends on both disk size and dust mass. The millimeter-wavelength spectral index can be significantly shallower than the naive expectation due to a combination of optical depth and deviations from the Rayleigh-Jeans regime. We fit radiative transfer models to the spectral energy distributions (SEDs) of 132 disks in the Taurus-Auriga region using a Markov chain Monte Carlo approach. We used all available data to produce the most complete SEDs used in any extant modeling study. We perform the fitting twice: first with unconstrained disk sizes and again imposing the disk size--brightness relation inferred for sources in Taurus. This constraint generally forces the disks to be smaller, warmer, and more optically thick. From both sets of fits, we find disks to be $\sim$1--5 times more massive than when derived using (sub-)millimeter measurements and common assumptions. With the uncertainties derived from our model fitting, the previously measured dust mass--stellar mass correlation is present in our study but only significant at the 2$\sigma$ level.
\end{abstract}


\section{INTRODUCTION}
\label{sec:introduction}

Protoplanetary disks of gas and dust around young stars are the birthplaces of planets. Measuring the properties of a representative sample of these disks is necessary to interpret the diversity of observed planetary systems and constrain models of planet formation.

The masses of protoplanetary disks are arguably their most important property with regard to the number and types of planets they may form. Absolute measurements of disk masses can be compared with the masses of known exoplanets \citep{najita2014} or the minimum-mass solar nebula \citep{weidenschilling1977,hayashi1981,desch2007} to assess whether the disks could form planets like those in our solar system. Measurements of disk masses relative to each other are also important for identifying correlations with other disk properties \citep[e.g.][]{tripathi2017,tazzari2017} and stellar properties \citep[e.g.][]{andrews2013,pascucci2016,ansdell2017,eisner2018}.

While dust is thought to comprise only $\sim$1\% of the protoplanetary disk mass (based on the dust fraction of the interstellar medium), it is the reservoir from which terrestrial planets and the cores of giant planets form. Thus, measurements of dust masses are crucial for assessing planet-forming potential. Furthermore, dust dominates the opacity of disks, meaning that observations of disks are more sensitive to the dust than to the gas component. Measurements of disk gas masses are subject to additional model-dependent complications, and disk gas-to-dust mass ratios remain uncertain \citep{williams2014,miotello2016,bergin2018}. In this study, we focus exclusively on measuring disk dust masses.

The dust mass of a disk is often calculated from its brightness in the (sub-)millimeter according to the analytic relation 
\begin{equation}
\label{eq:flux-mass}
M_\text{dust,ana} = \frac{F_\nu d^2}{\kappa B_\nu\left(T_\text{dust,ana}\right)}
\end{equation}
\citep{hildebrand1983,beckwith1990}. Here $F_\nu$ is the measured (sub-)millimeter flux density, $d$ is the distance to the disk from Earth, $\kappa$ is the dust opacity at the observed wavelength, and $B_\nu\left(T_\text{dust,ana}\right)$ is the Planck function at the average dust temperature. While computing dust masses with Equation \ref{eq:flux-mass} is common practice \citep[e.g.][]{andrews2005,andrews2013,carpenter2014,eisner2016,pascucci2016,eisner2018}, this method requires that specific values for the dust opacity and temperature be chosen. The relation is also predicated on the assumption that the disk is entirely optically thin to its own thermal emission at the observed wavelength. Throughout this paper, we will refer to the dust mass derived using Equation \ref{eq:flux-mass} as the ``analytic" mass.

The dust temperature used in Equation \ref{eq:flux-mass} (which we will refer to as the ``analytic" dust temperature) is sometimes taken to be 20 K. Another common approach is to scale the dust temperature with the luminosity of the host star, as in the relation
\begin{equation}
\label{eq:tdana}
T_\text{dust,ana} = 25 (L_\star/L_\sun)^{1/4} K
\end{equation}
used by \citet{andrews2013}. 

A more complete understanding of protoplanetary disks can be acquired by examining their spectral energy distributions (SEDs). The near-IR emission, or lack thereof, reveals the location of the disk's inner edge. The brightness in the mid-IR, where the disk is typically optically thick, traces the temperature of the disk surface. The shape of the SED from the mid- to far-IR indicates the vertical structure of the disk. The slope (spectral index) at long wavelengths can reveal the dust grain sizes, providing a more informed estimate of the dust opacity.

While first-order metrics of a disk SED---such as the spectral index between pairs of wavelengths---can serve as a basis for classification and relative comparison \citep[e.g.][]{lada1987}, a model that can reproduce the entire SED is preferable for relating the SED to the underlying physical properties of the disk. The simplest commonly employed model is a ``flat disk", where the dust surface density and temperature radial profiles are modeled as separate power laws. Flat-disk models, however, cannot constrain the disk's vertical structure, nor do they reflect the coupling between the disk structure and temperature. More sophisticated analytic models, such as the two-layer flared-disk model by \cite{chiang1997}, are also commonly used.

Radiative transfer modeling provides a more robust approach. In such models, photons from the central star are propagated into a specified dust distribution, defined on a cell-based grid. The temperature of the dust in each cell is computed by simulating the absorption and reemission of photons by the dust, and the simulated SED or image reflects the propagation of radiation out of the disk. This technique models the dust temperature and optical depth in a realistic manner, making it particularly useful for assessing the assumptions used in the analytic approaches. In Section \ref{sec:radiativetransfermodels}, we employ radiative transfer models to explore the effect of various disk parameters on the observable SEDs and properties of disks that are crucial for accurately measuring their dust masses (opacity, temperature, and optical depth).

Protoplanetary disks exhibit a diversity in their mass and other properties, and insights into planet formation can be made by exploring patterns in that diversity. To do so requires analyzing a large sample of disks with a coherent modeling framework. In Section \ref{sec:SEDmodeling}, we fit radiative transfer models to a large sample of disk SEDs in Taurus-Auriga. At $\sim$140 pc, Taurus is one of the nearest star forming regions. It is frequently targeted for observation, yielding well-sampled SEDs for most of its disk-bearing members. In Section \ref{sec:discussion}, we discuss the broader implications of our findings, and in Section \ref{sec:summary}, we summarize our results.

\section{RADIATIVE TRANSFER MODELS}
\label{sec:radiativetransfermodels}

\subsection{Disk Model Setup}
\label{sec:models}

Our disk model is azimuthally symmetric with a radial surface density profile following
\begin{equation}
\label{eq:sigma}
\Sigma(r)=\Sigma_0\left(\frac{r}{r_c}\right)^{-\gamma}\exp\left[-\left(\frac{r}{r_c}\right)^{2-\gamma}\right]
\end{equation}
from an inner edge $r_\text{in}$ to an outer edge $r_\text{out} = 10 \, r_c$, where $r_c$ is the characteristic disk size. Equation \ref{eq:sigma} is the profile predicted for a viscously accreting disk with viscosity varying as a power law with disk radius \citep{lynden-bell1974,hartmann1998}. We fixed the radial profile index to $\gamma$ = 1 for all models, as $\gamma \sim 1$ has been found by analyses of disks resolved at (sub-)millimeter wavelengths \citep{andrews2010,tazzari2016}. Varying $\gamma$ within reasonable bounds has a negligible effect on the SED \citep{woitke2016}, and independently constraining $\gamma$ requires spatially resolved disk observations. The normalization, $\Sigma_0$, is linked to the total dust mass as
\begin{equation}
\label{eq:sigma0}
\Sigma_0 = \frac{M_\text{dust}\left(\gamma-2\right)}{2\pi r_c^2 \left.\left[ \exp\left(-\left(\frac{r}{r_c}\right)^{2-\gamma}\right)\right]\right|_{r=r_\text{in}}^{r=r_\text{out}}}.
\end{equation}
When $r_\text{in} \ll r_c$ and $r_\text{out} \gg r_c$, $\Sigma_0 \approx M_\text{dust}\left(2-\gamma \right)/\left(2\pi r_c^2 \right)$. The volume density of dust follows
\begin{equation}
\label{eq:voldensity}
\rho(r,z) = \frac{\Sigma(r)}{\sqrt{2\pi} H(r)} \exp\left[-\frac{1}{2}\left(\frac{z}{H(r)}\right)^2\right]
\end{equation}
with scale height
\begin{equation}
H(r) = H_\text{100}\left(\frac{r}{\text{100 au}}\right)^\beta.
\end{equation}
Here $H_\text{100}$ sets the overall vertical extent of the disk, while $\beta$, the ``flaring parameter," determines how the scale height varies radially. Note that $r$ and $z$ in the preceding relations are cylindrical coordinates.

In theory, the disk vertical structure is set by hydrostatic equilibrium, so it could be determined self-consistently from the disk temperature profile computed by radiative transfer models. Indeed, some previous studies have adopted this approach \citep[e.g.][]{dullemond2004,mulders2012,hendler2017}. However, hydrostatic equilibrium only applies to the gas component, whereas radiative transfer calculations compute the dust temperature, so typically, the gas temperature is simply set equal to the dust temperature. Furthermore, the vertical distribution of the dust may differ from that of the gas due to, e.g., dust settling, which requires one or more additional free parameters to implement in the model. In practice, using hydrostatic equilibrium requires multiple iterations of radiative transfer calculations to find a self-consistent model, which makes the approach more computationally expensive. For these reasons, we opt to ignore the gas component and simply model the dust distribution directly.

Models of disk SEDs often use a power-law prescription for the spectral behavior of the dust opacity $\kappa (\lambda)$. While a single power law may be a good approximation of the real dust opacity at (sub-)millimeter wavelengths, it is less accurate at shorter wavelengths. Furthermore, independent parameters are often used to set the amplitude and slope of the power-law model. In reality, these parameters are correlated and determined by the more fundamental properties of the dust grains (e.g. sizes and compositions).

For our modeling, we computed the dust opacity, $\kappa (\lambda)$, with the DIANA Project Opacity Tool\footnote{http://dianaproject.wp.st-andrews.ac.uk/data-results-downloads/fortran-package/} \citep{woitke2016}. This code uses the optical constants of amorphous laboratory silicates \citep[Mg$_{0.7}$Fe$_{0.3}$SiO$_3$;][]{dorschner1995} and amorphous carbon \citep[BE-sample;][]{zubko1996}. It uses the distribution of hollow spheres method \citep{min2005}, for which we set the ``irregularity parameter" to the default value of $V_\text{hollow}^\text{max}$ = 0.8. We fixed the grain composition to the default mixture of 60\%  silicates, 15\% carbon, and 25\% porosity (vacuum). The amount of carbon in protoplanetary dust is not well constrained, but based on solar system estimates, the silicate/carbon ratio is often assumed to be roughly a few \citep{min2011}. A porosity fraction of 25\% has been shown to give good agreement with more realistic aggregate grain models \citep{min2016}. We do not explore the effect on the disk SEDs of varying the grain composition, but this has been investigated in other studies \citep[e.g.][]{miyake1993,dalessio2006,woitke2016}.

The grain sizes followed a power-law distribution $n(a) \propto a^{-q}$ (where $a$ is the grain radius) from $a_\text{min}$ to $a_\text{max}$ with $q$ and $a_\text{max}$ as free parameters and $a_\text{min}$ fixed to 0.05 $\micron$. We computed the opacities with 100 grain size bins at 300 wavelength points. The dust opacity was assumed to be constant throughout the disk. Spatial variations in grain properties are best studied with well-resolved images of disks at multiple radio wavelengths \citep[e.g.][]{perez2015,tazzari2016,tripathi2018}, rather than from the analysis of unresolved SEDs, as we conduct here.

We performed radiative transfer modeling with RADMC-3D \citep{dullemond2012} on a spherical coordinate grid. The radial grid spacing was divided into two regions in order to enhance the density of grid cells near the inner edge of the disk (as recommended by the RADMC-3D instruction manual), with the inner region from $r_\text{in}$ to 3 $r_\text{in}$ and the outer region from 3 $r_\text{in}$ to $r_\text{out}$. Each region had 60 radial grid steps distributed logarithmically. The grid spacing in polar angle $\theta$ was also divided into multiple regions to enhance the grid spacing near the disk midplane, with 10 grid steps from 0.1 to 1 rad, 80 grid steps from 1 to $\pi - 1$, and another 10 grid steps from $\pi - 1$ to $\pi - 0.1$. The grid spacing was uniform in each region. Because we assumed azimuthal symmetry, we used only two cells in azimuth. The dust density in each cell was determined from Equation \ref{eq:voldensity}, calculated at the center of the cell.

For the stellar spectrum (the central radiation source for the radiative transfer), we used the PHOENIX ``BT-settl" models \citep{baraffe2015} for stars $T_\star<$ 7000 K and ATLAS9 models \citep{castelli2004} when $T_\star>$ 7000 K.

Each radiative transfer model was performed in two steps. In the first step, the temperature of the dust was computed in each cell. We used $10^7$ photons at 200 wavelengths distributed logarithmically from 0.1 to 5000 $\micron$. We found that this many photons was necessary to maintain low noise in the temperature profile of the most dense disk models. We modeled the star as a spherical emitter, and we used the modified random walk algorithm. We did not include accretion heating in the disk, as this is typically only relevant in a small region of the overall disk. We also did not include heating from an external background radiation field, as previous radiative transfer studies of disks in this region have found this to have a negligible effect on the dust temperature \citep{vanderplas2016}. 
In the second step, the emission from the disk was simulated, yielding the model SED. We used $10^4$ photons to model the SED at 200 wavelengths from 0.1 to 5000 $\micron$. Scattered light from the dust in the disk was not included in the model SED, but the direct contribution of flux from the star was included.

\subsection{Exploration of Model Parameters}
\label{sec:parameterdemo}

\begin{figure*}
\epsscale{1.15}
\plotone{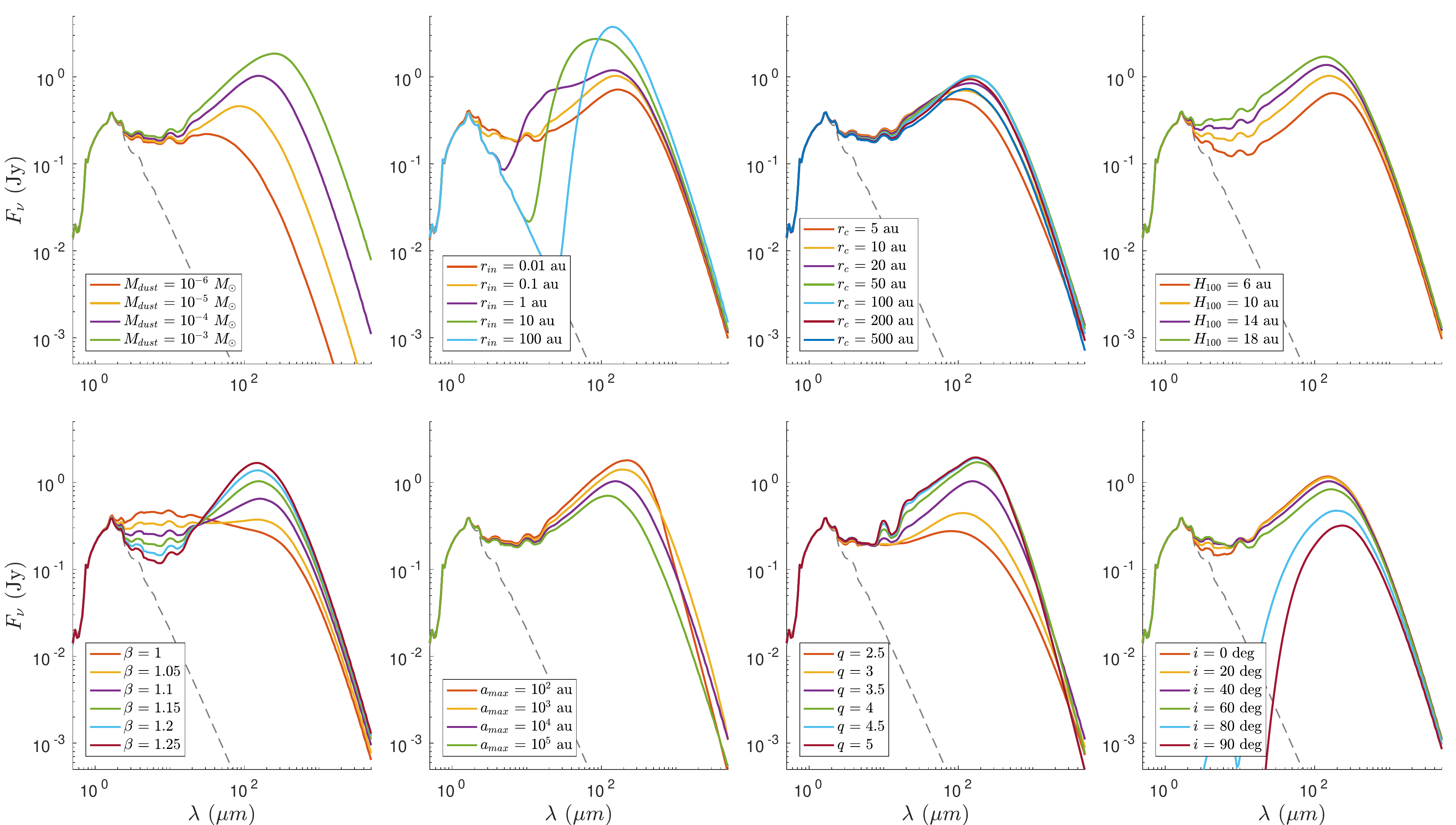}
\caption{Demonstration of the effect on the model SED by varying each of the eight free parameters from the fiducial model. The dashed gray line in each plot is the stellar photosphere.}
\label{fig:paramdemoSEDs}
\end{figure*}

\begin{deluxetable*}{lclllc}
\tabletypesize{\scriptsize}
\tablewidth{0pt}
\tablecolumns{6}
\tablecaption{Disk Model Free Parameters \label{table:parameters}}
\tablehead{\colhead{Parameter} & \colhead{Symbol} & \colhead{Fiducial} & \colhead{Allowed} & \colhead{Initialized} & \colhead{Linear or} \\ \colhead{} & \colhead{} & \colhead{Value} & \colhead{Values} & \colhead{Values} & \colhead{Logarithmic}}
\startdata
Dust mass & $M_\text{dust}$ & $10^{-4} M_\sun$ & $(0 \cdots \infty)$ & $(10^{-6} \cdots 10^{-2.5})\,M_\sun$ & Log \\ 
Inner edge & $r_\text{in}$ & 0.1 au & $(1.01R_\star \cdots r_c)$ & $(10^{-2} \cdots 10^{2})$ au & Log \\ 
Characteristic size & $r_c$ & 100 au & $(3 \cdots 10^4)$ au & $(10^{0.8} \cdots 10^{3.5})$ au & Log \\ 
Scale height at 100 au & $H_\text{100}$ & 10 au & $(0 \cdots \infty)$ & $(3 \cdots 20)$ au & Linear \\
Flaring parameter & $\beta$ & 1.15 & $(1 \cdots \infty)$ & $(1 \cdots 1.25)$ & Linear \\
Maximum grain size & $a_\text{max}$ & $10^4$ $\micron$ & $(0.05 \,\micron \cdots \infty)$ & $(10^{1} \cdots 10^{6})\,\micron $ & Log \\
Grain size distribution index & $q$ & 3.5 & $(0 \cdots \infty)$ & $(1 \cdots 5)$ & Linear \\
Inclination & $i$ & 40$^\circ$ & $(0^\circ \cdots 90^\circ)$ & $(0^\circ \cdots 90^\circ)$ & Linear \\
\enddata
\end{deluxetable*}

To explore the effects of the model parameters, we constructed a fiducial model and varied each parameter individually from its fiducial value. The eight free parameters are summarized in Table \ref{table:parameters}. The fiducial model has $M_\text{dust}$ = $10^{-4} M_\sun$, $r_\text{in}$ = 0.1 au, $r_c$ = 100 au, $H_\text{100}$ = 10 au, $\beta$ = 1.15, $a_\text{max}$ = $10^{4}$ $\micron$, $q$ = 3.5, and $i$ = 40$^\circ$. We fixed the stellar parameters to $T_\star$ = 3500 K and $L_\star$ = 0.5 $L_\sun$.

The effect on the model SED is presented in Figure \ref{fig:paramdemoSEDs}. Here $M_\text{dust}$ has little effect on the SED at short wavelengths where the disk is optically thick, but the flux density at long wavelengths scales roughly linearly with $M_\text{dust}$, making (sub-)millimeter observations crucial to measuring dust masses. Increasing $r_\text{in}$ depletes the disk of hot and warm dust near the star, thus reducing the near-IR and then mid-IR emission from the disk. An SED lacking in short-wavelength excess emission is the characteristic signature of a transition disk with a cleared inner cavity. Compared with other parameters, $r_c$ has a smaller influence on the overall SED, with only very small disks having a noticeable effect. Spatially resolved observations of disks are usually required to constrain the disk size. Increasing the scale height, $H_\text{100}$, increases the amount of the stellar radiation absorbed by the surface of the disk, increasing its temperature and thus its brightness in the infrared. A highly flared disk will intercept more flux in its outer part and less in its inner regions, making the disk brighter in the far-IR and fainter in the near-IR. A flatter (less flared) disk exhibits the opposite behavior. The disk inclination has little effect on the SED except when it is close to edge-on, in which case the central star and inner hot regions of the disk are occulted. The influence of the disk structure on the SED presented here generally agrees with the results from similar studies \citep[e.g.][]{miyake1995,dalessio1999,woitke2016}.      

\begin{figure}
\epsscale{1}
\plotone{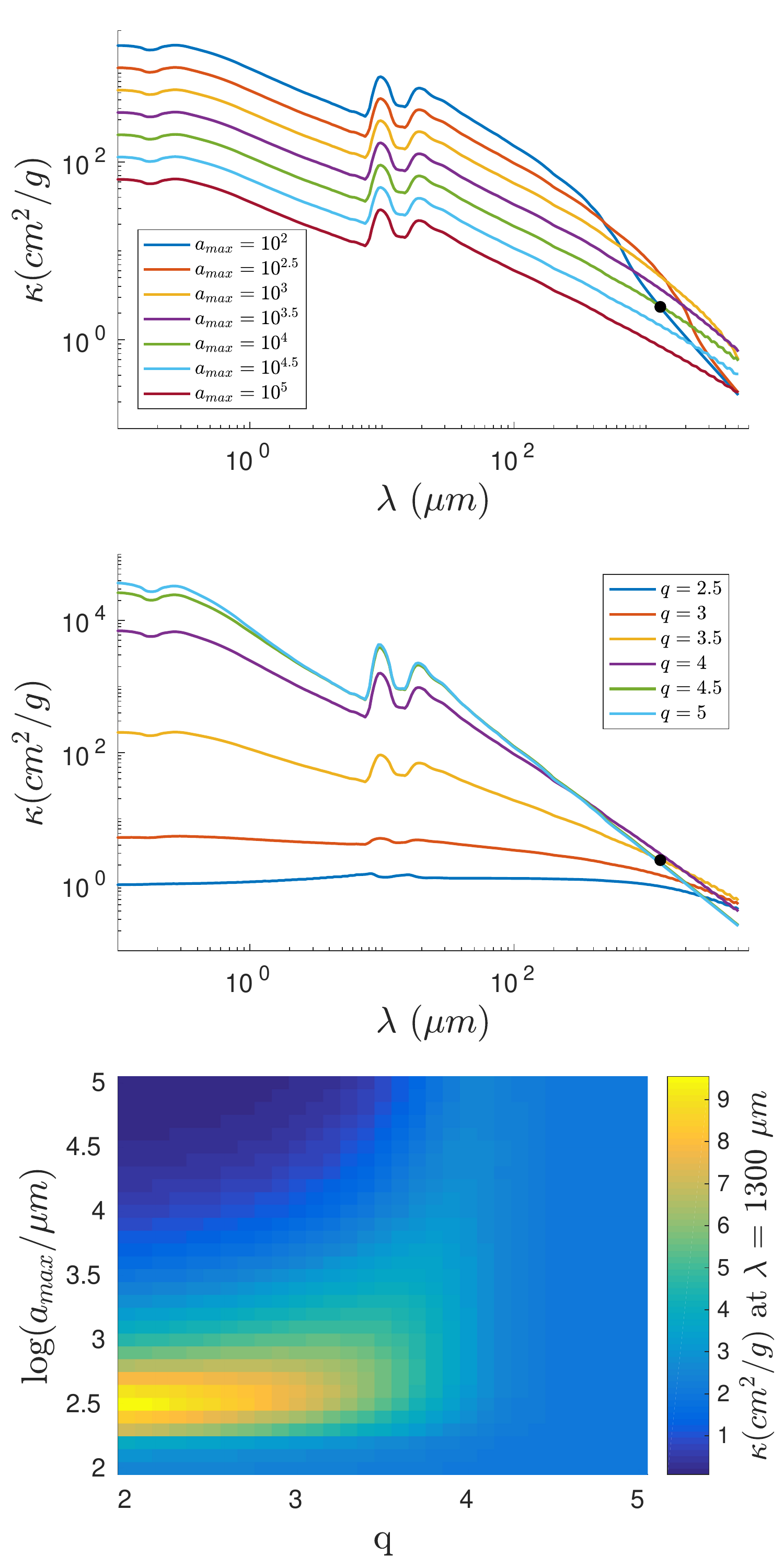}
\caption{Top---Effect of varying the maximum grain size, $a_\text{max}$, on the dust opacity spectrum with the index of the size distribution, $q$, fixed at 3.5. Middle---Same as above but varying $q$ and fixing $a_\text{max}$ to $10^{4}$ $\micron$. In both plots, $\kappa$ = 2.3 g/cm$^2$ at a wavelength of 1300 $\micron$ is indicated with a black point. This is the value commonly assumed by previous studies. Bottom---Effect of jointly varying $a_\text{max}$ and $q$ on the opacity value at 1300 $\micron$.}
\label{fig:paramdemokappa}
\end{figure}

The dust grain properties ($a_\text{max}$ and $q$) influence the observed disk SED via the opacity spectrum, which we show in more detail in Figure \ref{fig:paramdemokappa}. Smaller $a_\text{max}$ values shift the entire opacity spectrum higher. They also result in a steeper slope in the (sub-)millimeter regime starting at shorter wavelengths, which translates to a steeper spectral index in the SED. Higher values of $q$ (a steeper grain size distribution and thus more small grains versus large grains) also lead to higher opacity over much of the spectrum and a steeper slope at long wavelengths. The effects of $a_\text{max}$ and $q$ on the opacity spectrum agree, in general, with calculations performed using Mie theory \citep{miyake1993}. In the bottom panel of Figure \ref{fig:paramdemokappa}, we show the effect on the opacity at 1300 $\micron$ (a common wavelength at which disks are observed and their \text{dust} masses calculated) by jointly varying $a_\text{max}$ and $q$. In this fairly broad range of parameter space, the opacity differed from the commonly assumed value of 2.3 cm$^2$/g by up to a factor of $\sim$4.           

\begin{figure*}
\epsscale{1.15}
\plotone{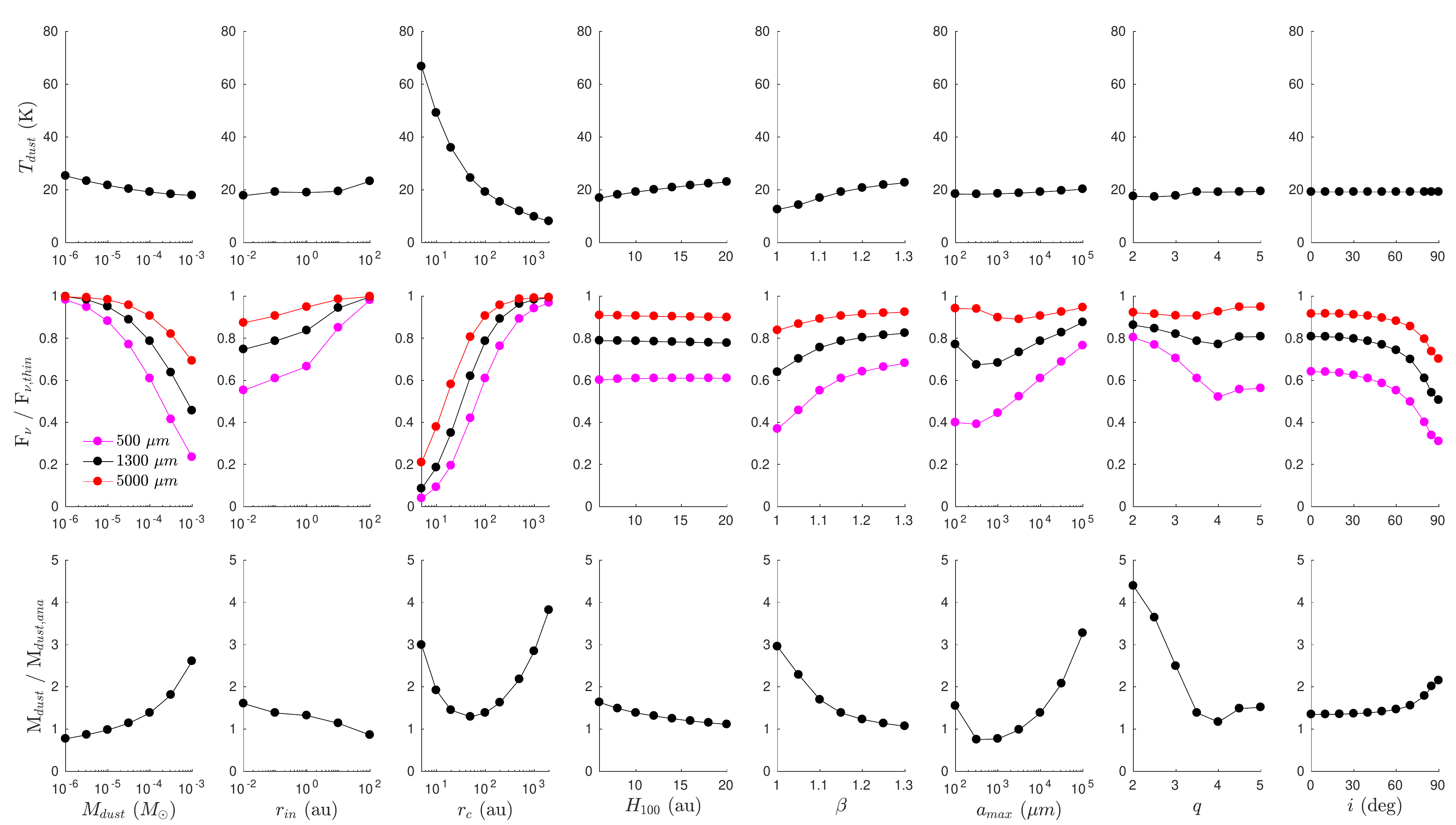}
\caption{Effect of each of our eight disk free parameters (varied individually from the fiducial model) on the average dust temperature ($T_\text{dust}$; top row), $F_\nu/F_{\nu,\text{thin}}$ at three different wavelengths (middle row), and $M_\text{dust}/M_\text{dust,ana}$ (bottom row). Here $T_\text{dust}$ is computed from Equation \ref{eq:Td}, $F_{\nu,\text{thin}}$ is computed from Equation \ref{eq:Fthin}, and $M_\text{dust,ana}$ is computed from Equation \ref{eq:flux-mass}.}
\label{fig:paramdemoallvsall}
\end{figure*}

The effect on the dust temperature of each parameter (with the other parameters fixed to their fiducial values) is shown in the top row of Figure \ref{fig:paramdemoallvsall}. This is the mass-weighted average dust temperature, 
\begin{equation}
\label{eq:Td}
T_\text{dust} = \frac{1}{M_\text{dust}} \sum_i^{N_{\text{cells}}} m_{\text{dust},i} T_{\text{dust},i},
\end{equation}
with $m_{\text{dust},i}$ the mass of dust in each cell and $T_{\text{dust},i}$ the temperature of the dust in each cell. We find that the size of the disk ($r_c$) has by far the greatest effect on the disk temperature. Smaller disks can be significantly warmer than is typically assumed for $T_\text{dust,ana}$, in agreement with expectations and the findings of \citet{hendler2017}. Note that for this fiducial model, $T_\text{dust,ana} \approx$ 20 K (from Equation \ref{eq:tdana} with $L_\star$ = 0.5 $L_\sun$).   

We also tested the effect of the model parameters on the optical depth of the disk to its thermal emission. Our primary interest here was to assess the effect of optical depth on the translation from disk flux in the (sub-)millimeter to dust mass (i.e. the assumption inherent in Equation \ref{eq:flux-mass} that the disk is optically thin). Thus, we quantify the optical depth with the metric $F_\nu / F_{\nu,\text{thin}}$, where $F_\nu$ is the flux returned by the radiative transfer model and $F_{\nu,\text{thin}}$ is the flux that would be expected if the disk were perfectly optically thin to its own thermal emission. We compute the latter as
\begin{equation}
\label{eq:Fthin}
F_{\nu,\text{thin}} = \frac{\kappa}{d^2} \sum_i^{N_{\text{cells}}} m_{\text{dust},i} B_\nu(T_{\text{dust},i}).
\end{equation}
Equation \ref{eq:Fthin} is the sum of the expected flux from each cell in the model grid. The opacity and distance have been factored out of the sum because they do not vary from cell to cell. When $B_\nu(T_\text{dust}) \propto T_\text{dust}$ (as in the Rayleigh-Jeans regime), $F_{\nu,\text{thin}}$ is equivalent to Equation \ref{eq:flux-mass} (solved for $F_\nu$) using the mass-weighted average $T_\text{dust}$.

In the middle row of Figure \ref{fig:paramdemoallvsall} we show $F_\nu / F_{\nu,\text{thin}}$ at 500, 1300, and 5000 $\micron$, and we find, as expected, that the optical depth is lower at longer wavelengths. Of the eight free parameters, higher dust masses and smaller disk sizes lead to the greatest increases in optical depth (lower $F_\nu/F_{\nu,\text{thin}}$) because they result in a higher surface density. Other parameters that increase the optical depth include smaller $r_\text{in}$ (more mass in the dense inner region of the disk), lower $\beta$ (a more vertically compact disk), higher inclinations (more mass along the line of sight), and dust properties ($a_\text{max}$ and $q$) that yield a higher $\kappa$.

In the bottom row of Figure \ref{fig:paramdemoallvsall} we show the effect of each parameter on the ratio of $M_\text{dust}$ (the true mass of the dust in the model) to $M_\text{dust,ana}$ (the mass derived from the flux density of the model disk at 1300 $\micron$ using equations \ref{eq:flux-mass} and \ref{eq:tdana} and $\kappa$ = 2.3 cm$^2$/g). Dust properties yielding low values of $\kappa$ lead to higher $M_\text{dust}/M_\text{dust,ana}$. Disk properties that increase the optical depth also result in higher $M_\text{dust}/M_\text{dust,ana}$, namely, higher $M_\text{dust}$, low $\beta$, and high inclinations. The disk size ($r_c$) has a strong influence on both the disk temperature and optical depth, resulting in a more complicated effect on $M_\text{dust}/M_\text{dust,ana}$. Large disks are optically thin and colder than $T_\text{dust,ana}$, leading to a higher $M_\text{dust}/M_\text{dust,ana}$. Medium-sized disks have $T_\text{dust}$ $\approx$ $T_\text{dust,ana}$ (20 K), but they are slightly optically thick, and thus $M_\text{dust}/M_\text{dust,ana}$ is above unity. Small disks have hotter dust and higher optical depths---effects that act in opposing directions on $M_\text{dust}/M_\text{dust,ana}$---but the optical depth effect is stronger, so $M_\text{dust}/M_\text{dust,ana}$ increases.

\begin{figure}
\epsscale{1}
\plotone{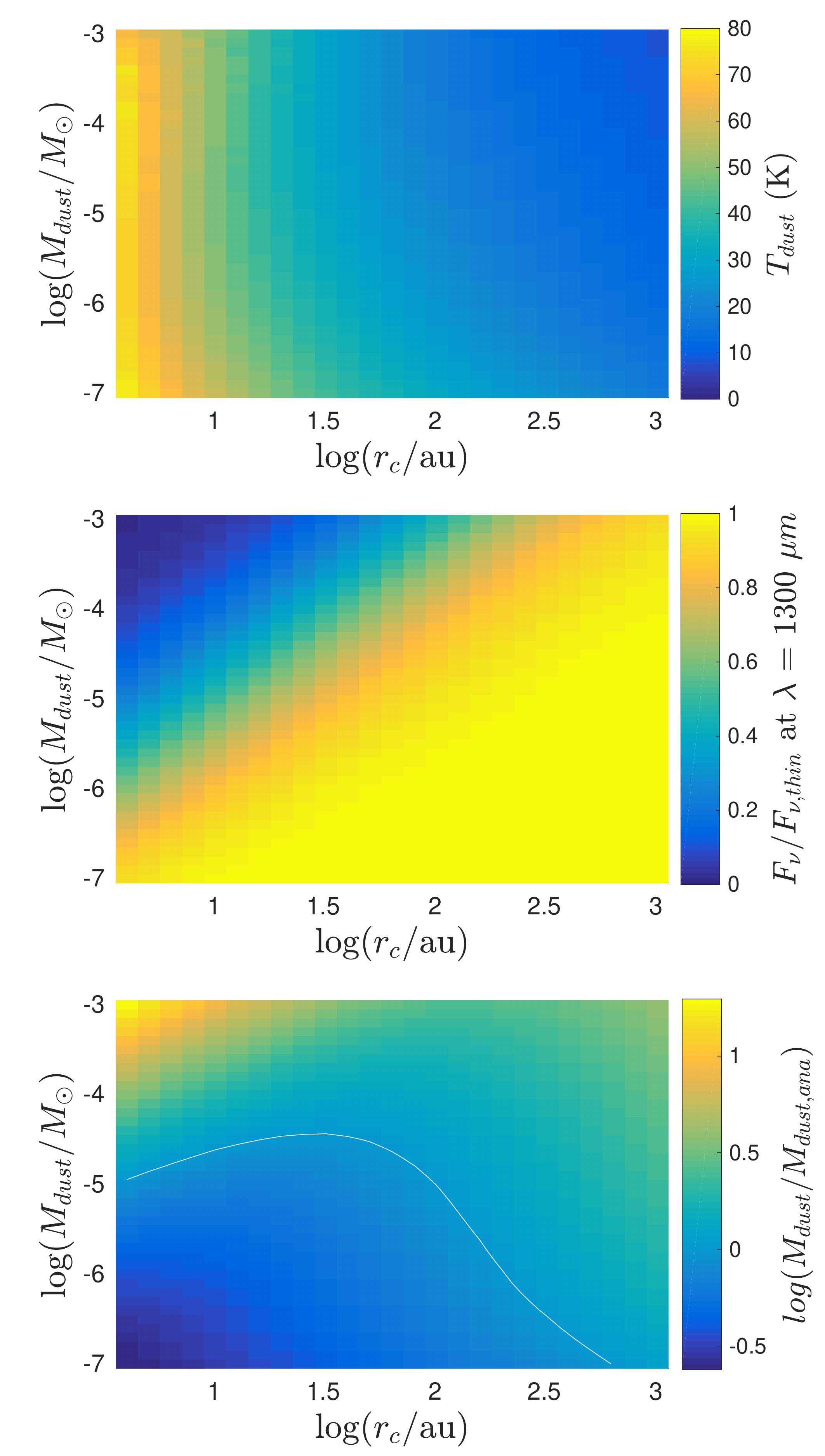}
\caption{Effect of jointly varying the dust mass and disk size while holding other disk parameters fixed to the fiducial values on the average dust temperature ($T_\text{dust}$; top), the optical depth (as described by $F_\nu/F_{\nu,\text{thin}}$ at 1300 $\micron$; middle), and $M_\text{dust}/M_\text{dust,ana}$ (bottom). The white line in the bottom panel marks where the ratio is unity.}
\label{fig:paramdemosizemassjoint}
\end{figure}

\begin{figure*}
\epsscale{1.15}
\plotone{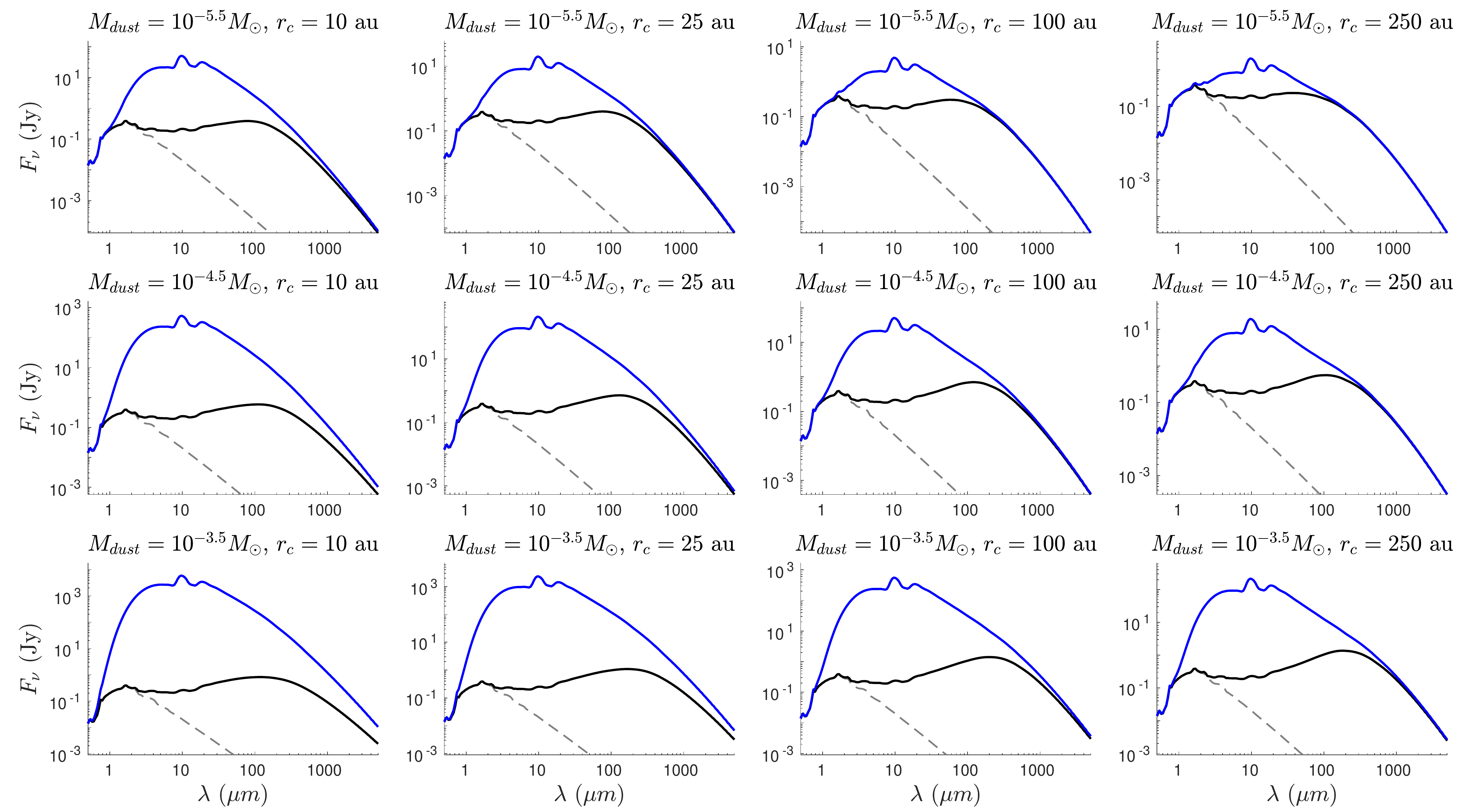}
\caption{The $F_\nu(\lambda)$ returned from the full radiative transfer model (black) compared with $F_{\nu,thin}(\lambda)$ added to the flux from the stellar photosphere (blue) for disks with a variety of dust masses (rows) and disk sizes (columns). The gray dashed line in each plot is the stellar photosphere.}
\label{fig:paramdemothinseds}
\end{figure*}

\subsection{Disk Size and Dust Mass}

Since the disk size and dust mass both have significant effects on the temperature and optical depth, we next explored varying them jointly (while maintaining the other six parameters at their fiducial values). We show the results in Figure \ref{fig:paramdemosizemassjoint}. The top panel again shows that the disk temperature depends primarily on disk size (although less massive disks of the same size are also slightly warmer), and that small disks are significantly warmer than assumed for $T_\text{dust,ana}$. The center panel illustrates that small and/or massive disks are not entirely optically thin, even at a wavelength of 1300 $\micron$, where Equation \ref{eq:flux-mass} is often used. The bottom panel illustrates that $M_\text{dust}$ can, in principle, be greater or less than $M_\text{dust,ana}$, depending on the particular dust mass and disk size. The white line shows the locus where $M_\text{dust}$ = $M_\text{dust,ana}$. In the lower right region of the plot, this locus indicates where these masses agree for the ``right" reasons---that is, the disk is truly optically thin and $T_\text{dust,ana}$ is an accurate measure of the average dust temperature. On the left side of the plot, however, the disks are not optically thin, so the dust masses agree only when the temperature is such that it counterbalances the optical depth effects.

In Figure \ref{fig:paramdemothinseds} we show SEDs for both $F_{\nu,\text{thin}}(\lambda)$ and the full radiative transfer models for disks with a range of dust masses and disk sizes. This illustrates that the wavelength at which disks become optically thin can vary significantly, depending on these properties. For instance, a large and low-mass disk becomes optically thin in the far-IR, so Equation \ref{eq:flux-mass} could be applied relatively accurately to compute the disk mass from \textit{Herschel}/SPIRE photometry, rather than requiring a (sub-)millimeter detection. On the other hand, a small and massive disk may not be entirely optically thin even at $\lambda$ $\sim$ 5 mm, wavelengths where the optically thin assumption is usually not questioned.

\begin{figure*}
\epsscale{1}
\plotone{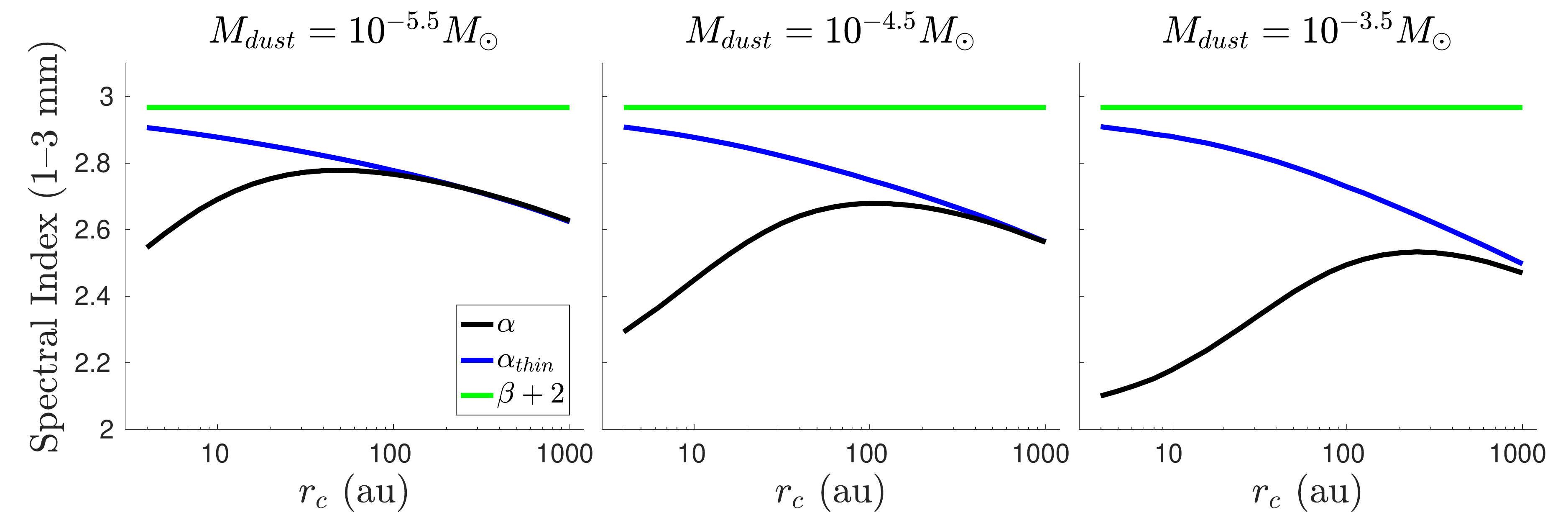}
\caption{Spectral index ($\alpha$) measured between wavelengths of 1 and 3 mm vs. disk size for three different model dust masses. The spectral indices of the full radiative transfer models are shown in black, and those of the optically thin models are shown in blue. These are compared with $\beta$ + 2 (green), the expectation for $\alpha$ in the case of a completely optically thin disk in the Rayleigh-Jeans regime. The difference between the green and blue lines is due to deviations from the Rayleigh-Jeans regime, and the difference between the blue and black lines is due to optical depth effects.}
\label{fig:paramdemospectralindex}
\end{figure*}

\subsection{The Spectral Index}

Also apparent from Figure \ref{fig:paramdemothinseds} is that for the model disks that are not entirely optically thin in the (sub-)millimeter, the slope (spectral index) at these wavelengths is shallower than that of $F_{\nu,\text{thin}}(\lambda)$. Measuring the spectral index is important for constraining the maximum grain size to study the process of grain growth and to accurately calculate the dust mass \citep[e.g.][]{beckwith1991,testi2014,ribas2017}. The spectral index can also vary with the dust composition \citep{pollack1994,dalessio2001}, although we do not explore that dependence here.

In the (sub-)millimeter regime, the opacity spectrum is often approximated as a power law $\kappa(\lambda) \propto \lambda^{-\beta}$. (Note that the use of the variable $\beta$ here has no relation to the disk flaring parameter.) In the optically thin case, $F_{\nu,\text{thin}}(\lambda) \propto B_\nu(\lambda) \kappa(\lambda)$. Further assuming that the disk emission is in the Rayleigh-Jeans regime yields $F_{\nu,\text{thin}}(\lambda) \propto \lambda^{-(\beta+2)}$. Thus, from measuring the spectral index ($\alpha$), the slope of the opacity spectrum can be calculated as $\beta = \alpha-2$.

We used our radiative transfer models to investigate the accuracy of this method for a range of dust masses and disk sizes, and the results are shown in Figure \ref{fig:paramdemospectralindex}. For the fiducial model dust properties, $\beta \approx$ 1 (measured between wavelengths of 1 and 3 mm), so if these assumptions hold, we would expect $\alpha \approx$ 3. We find that both the full radiative transfer model, $F_\nu(\lambda)$, and the optically thin model, $F_{\nu,\text{thin}}(\lambda)$, have lower (shallower) spectral indices than this ideal expectation. For the optically thin case, the shallower spectral index is a result of the disk emission not being perfectly in the Rayleigh-Jeans regime. The discrepancy is minimized for smaller---and thus warmer---disks for which the Rayleigh-Jeans approximation is more accurate. The additional discrepancy between the optically thin spectral index and that of the full radiative transfer model is due to optical depth effects. As expected, this discrepancy is more pronounced for smaller and more massive disks. The fact that optical depth tends to decrease the spectral index can be understood by considering the limiting case of a completely optically thick disk, for which $F_{\nu,\text{thick}}(\lambda) \propto B_\nu(\lambda)$, and thus $\alpha \lesssim$ 2. The complications that arise when interpreting the spectral index due to temperature and optical depth effects have been discussed in the literature \citep[e.g.][]{beckwith1990,testi2001,andrews2005,ricci2012}. We find that radiaitve transfer models provide a valuable tool to isolate and quantify these effects.

In this section, we illustrated some of the complications involved when retrieving fundamental disk properties from observed SEDs. Fortunately, these complications can be accounted for when interpreting observations by fitting them with radiative transfer models. In Section \ref{sec:SEDmodeling}, we fit models to the observed SEDs of disks in the Taurus-Auriga star-forming region. 

\section{MODELING TAURUS DISKS}
\label{sec:SEDmodeling}

\subsection{Target Selection and Data}
\label{sec:targets}

We adopted the sample of class II sources in Taurus from \citet{andrews2013}, which they argued was fairly complete. This totaled 178 systems. For the stellar properties ($T_\star$, $L_\star$), we used the best-fit values from Table 4 of \cite{andrews2013}.

We discarded systems from our sample that were (1) edge-on disks or (2) disks in close binary or multiple systems. Edge-on disks obscure the star to some degree, so the inferred disk and stellar properties become correlated. In our fitting procedure, however, we keep the stellar properties fixed for a given source. Furthermore, a nonnegligible contribution to the optical and near-IR flux of edge-on systems may come from scattered light \citep[e.g.][]{luhman2007}, which is not included in our models. Some of the systems we discarded may actually be class I (embedded) sources, which can be mistaken for edge-on class II disks, but in either case, removal from the sample is appropriate. 

We discarded close binary/multiple systems for which disk emission could not reliably be attributed to a specific star. In these cases, \citet{andrews2013} assigned (sub-)millimeter detections to the primary components and upper limits on the flux to the secondaries. We do not adopt this procedure, as ALMA observations by \citet{akeson2014} have found that circumsecondary disks can be more massive than circumprimary disks. Only in cases where high-resolution observations (that resolve the components) show only one star hosting a significant disk do we assign the full measured SED to that star. In cases where multiple components host disks that are resolved at some wavelengths but not others, we simply exclude the confused data points from our fitting. We do include a few cases of known circumbinary disks, for which we modeled the central star with $L_\star = L_{\star,A} + L_{\star,B}$ and $T_\star = (T_{\star,A} L_{\star,A} + T_{\star,B} L_{\star,B}) / (L_{\star,A} + L_{\star,B})$. We fit models to \NumFinalTargets{} disks from the original sample of 178. Notes on specific systems---including those that we discarded from the \cite{andrews2013} sample---are given in the Appendix.  

We used the photometry for each target provided by \cite{andrews2013}. To this, we added additional measurements from the literature, primarily at far-IR and (sub-)millimeter wavelengths. These new data are listed in Table \ref{table:data}. Our compiled photometry is the most complete set of SED data yet assembled for class II sources in Taurus. We dereddened the data using $A_V$ values for each target from Table 4 of \citet{andrews2013} and extinction curves from \citet{mcclure2009}. Total uncertainties were computed from the combination in quadrature of the statistical and calibration uncertainties. To ensure that no data point was weighted too highly in our fitting, any point with a final uncertainty of $<$ 5\% was scaled up to 5\%. We excluded some data points from the fitting, typically those at UV-to-visible wavelengths that showed an excess above the photosphere model. This excess luminosity is likely due to accretion, which is not accounted for in our models. The reasons for excluding other measurements are described for individual targets in the Appendix.

\begin{deluxetable*}{ccccccc}
\tabletypesize{\scriptsize}
\tablewidth{0pt}
\tablecolumns{7}
\tablecaption{Additional SED Data \label{table:data}}
\tablehead{\colhead{Target} & \colhead{$\lambda$} & \colhead{$F_\nu$} & \colhead{Stat $\sigma$} & \colhead{Cal $\sigma$} & \colhead{Instrument} & \colhead{References} \\ \colhead{} & \colhead{($\micron$)} & \colhead{(mJy)} & \colhead{(mJy)} & \colhead{(\%)} & \colhead{} & \colhead{}}
\startdata
AA Tau & 70 & 1172.60 & 37.00 & 5 & \textit{Herschel}/PACS & PACS Point Source Catalog \\
AA Tau & 100 & 1031.00 & 33.60 & 5 & \textit{Herschel}/PACS & PACS Point Source Catalog \\
AA Tau & 160 & 1213.40 & 76.10 & 5 & \textit{Herschel}/PACS & PACS Point Source Catalog \\
AA Tau & 250 & 1103.50 & 38.00 & 4 & \textit{Herschel}/SPIRE & SPIRE Point Source Catalog \\
AA Tau & 350 & 917.90 & 44.10 & 4 & \textit{Herschel}/SPIRE & SPIRE Point Source Catalog \\
AA Tau & 500 & 624.10 & 38.10 & 4 & \textit{Herschel}/SPIRE & SPIRE Point Source Catalog \\
AA Tau & 1060 & 106.10 & 0.54 & 20 & ALMA & \citet{loomis2017} \\
AA Tau & 1150 & 86.60 & 0.53 & 20 & ALMA & \citet{loomis2017} \\
AA Tau & 1300 & 54.80 & 0.70 & 20 & SMA & \citet{williams2014} \\
\enddata
\tablecomments{See \citet{marton2017} for details regarding the PACS Point Source Catalog and \citet{schulz2017} for details regarding the SPIRE Point Source Catalog. (This table is available in its entirety in machine-readable form.)}
\end{deluxetable*}

\subsection{SED Fitting Procedure}
\label{sec:fitting}

We fit models to the SED of each target using the Markov chain Monte Carlo (MCMC) software package emcee \citep{foreman-mackey2013}. We ran 300 walkers in parallel for each target. The model generation and fitting was run on the University of Arizona High Performance Computing system.

We varied the eight free parameters introduced in Section \ref{sec:radiativetransfermodels}. For parameters that were likely to vary by more than an order of magnitude, we used the base-10 $\log$ of the parameter in the fitting. The starting parameters for each walker were selected randomly from a uniform distribution, as noted in Table \ref{table:parameters}. 

The natural log of the likelihood (the goodness-of-fit metric used by emcee) was computed as
\begin{eqnarray}
\label{eq:lnlike}
\ln P = -\frac{1}{2}\sum_i^{N_{\text{det}}}\left(\frac{\log F_{\nu,\text{model},i} - \log F_{\nu,\text{det},i}}{\sigma_{\text{det},i}/(F_{\nu,\text{det},i}\ln10)}\right)^2 \nonumber\\
+\sum_j^{N_{\text{ul}}}\ln \left[ \frac{1}{2} \, \text{erfc} \left(\frac{\log F_{\nu,\text{model},j}- \log F_{\nu,\text{ul},j}}{\sqrt{2} \, \log F_{\nu,\text{ul},j}}\right)  \right],
\end{eqnarray}
where $F_{\nu,\text{model}}$ are the model flux density values interpolated onto the observed wavelengths, and erfc is the complementary error function. 
The first sum in Equation \ref{eq:lnlike} is over the detected flux density measurements and equivalent to the usual $\chi^2$ metric. The second sum is over the nondetection measurements ($F_{\nu,\text{ul},j}$ is the 1$\sigma$ upper limit) and based on the formalism by \citet{sawicki2012}. The fit, as indicated in Equation \ref{eq:lnlike}, was performed on the $\log$ of the flux densities. We found that, in practice, Equation \ref{eq:lnlike} is weighted toward fitting the detections, with a comparatively small penalty induced when models violated upper-limit measurements. For some systems, this turned out to be beneficial, as there were measured detections at higher flux densities than upper limits reported at the same wavelengths, and it was our preference that the fits should prioritize the detections. 

Parameters were bounded to the ranges given in Table \ref{table:parameters}. When emcee selected a model with one or more parameters outside of these ranges, the radiative transfer was not performed, and $\ln P$ was set to $-\infty$. We chose these bounds to be as nonrestrictive as possible, i.e. not influenced by prior modeling of disks or theoretical predictions. We examined the distributions of the model parameters versus the steps of the MCMC to identify where the results converged, and we discarded all previous steps in the chains. The remaining steps defined the posterior set of models used for our analysis.

\subsection{Disk Size Constraint}

In section \ref{sec:radiativetransfermodels}, we showed that the disk size has a large influence on the dust temperature and optical depth and thus on the derived dust mass. However, we also showed in Figure \ref{fig:paramdemoSEDs} that the disk size has only a minor effect on the observed SED and thus is difficult to constrain from fits to SEDs alone.

With this in mind, we performed our fitting of each target again, the second time with an external constraint on the disk size imposed. This constraint was the relation between disk size and submillimeter brightness established by \citet{tripathi2017} from spatially resolved observations of disks at 880 $\micron$, specifically,      
\begin{equation}
\label{eq:sizeconstraint}
\log R_\text{eff} = (2.12 \pm 0.05) + (0.5 \pm 0.07) \log F_\nu.
\end{equation}
Here $R_\text{eff}$ is the radius of the (inclination-deprojected) disk image that encompasses 68\% of the total flux density at 880 $\micron$. The relation was derived using disks from Taurus and Ophiuchus, but many of our targets were not included, so we could not adopt individual disk sizes directly. 

For each model SED generated during our fitting process, we also generated a model disk image\footnote{The model image was made with RADMC-3D on a 512 $\times$ 512 pixel grid using $10^5$ photons.} at 880 $\micron$ from which we computed $R_\text{eff}$. We compared this with the $R_\text{eff}$ expected from equation \ref{eq:sizeconstraint} based on the model total flux density at 880 $\micron$, and we computed the $\chi^2$ metric with the uncertainty factor calculated from combining in quadrature the uncertainties on the slope and intercept of equation \ref{eq:sizeconstraint} with 0.2 dex of scatter \citep{tripathi2017}. We then added this $\chi^2$ as an extra term to the fit metric given by Equation \ref{eq:lnlike}.

However, this alone would have resulted in the disk size constraint influencing $\ln P$ with approximately the same weight as a single photometric data point in the SED fit, so we opted to increase its weight to be of equal order to the entire SED fit. We examined the magnitudes of the size constraint $\chi^2$ terms compared with the SED fit $\ln P$ terms in the posterior set of models from the initial fitting (without the size constraint included in the fitting). The factor by which to increase the weighting was the ratio of the median of the latter to the median of the former. This extra weighting factor was determined for each target individually and then applied when running the second fit (with the size constraint included).

\begin{figure}
\epsscale{1.15}
\plotone{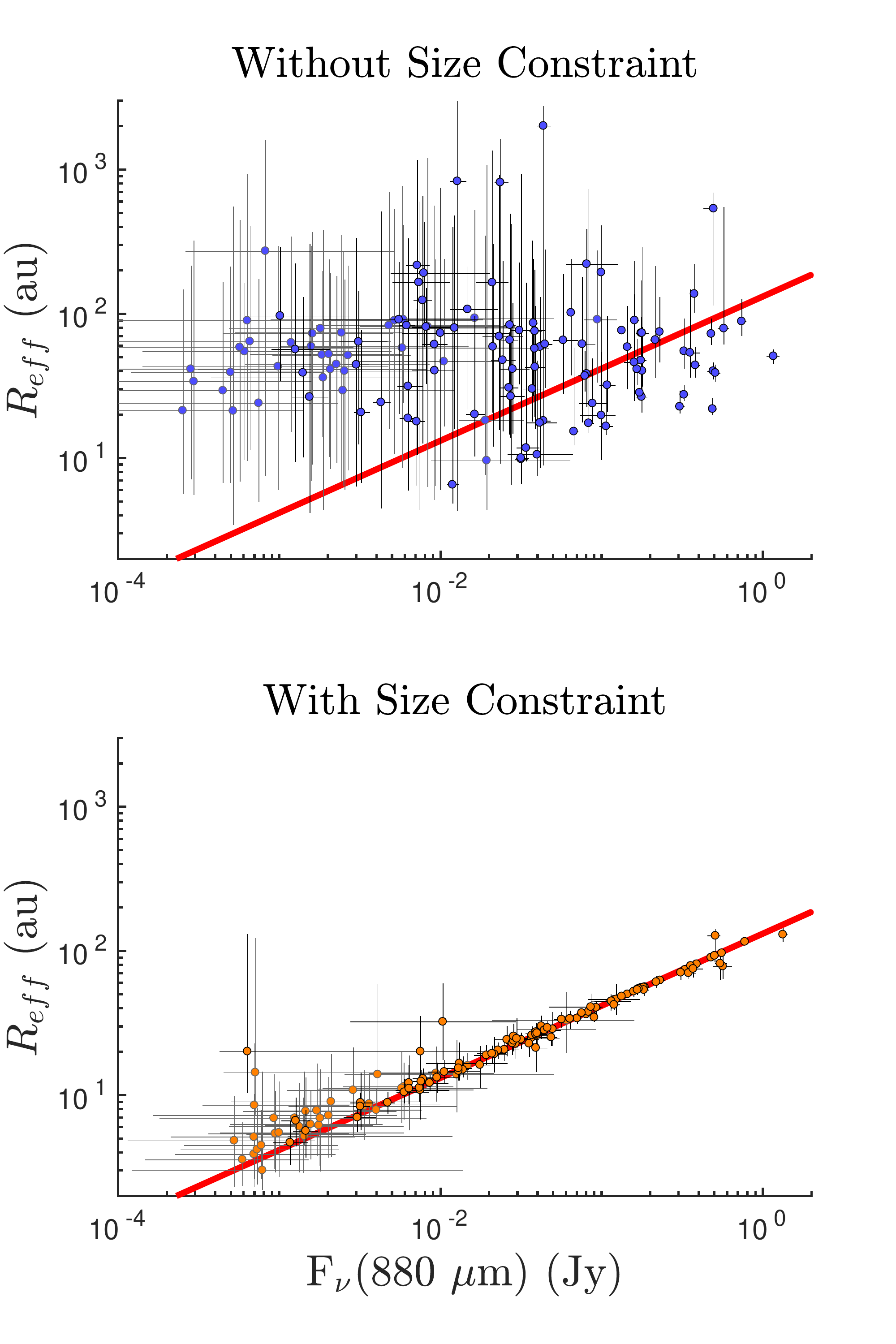}
\caption{Effect of implementing the external size constraint in the fitting. Both panels show $R_\text{eff}$ vs. $F_\nu$ at 880 $\micron$, with the red line depicting the constraint (Equation \ref{eq:sizeconstraint}). The top panel shows the results of the fitting without the constraint included, and bottom shows those with the constraint. The points outlined in black and with black error bars are systems with detections at $\lambda >$ 500 $\micron$, while those in gray have no measurements or only upper limits at $\lambda >$ 500 $\micron$.}
\label{fig:comparewithprior}
\end{figure}

In Figure \ref{fig:comparewithprior} we illustrate the effect of including the size constraint in the fitting. Both panels plot $R_\text{eff}$ versus $F_\nu$ at 880 $\micron$, and the red line shows the size constraint itself (Equation \ref{eq:sizeconstraint}). The top panel shows the unconstrained results for reference. The majority of disks tend to be larger than the external constraint would predict, and the range of disk sizes in the posterior set of models for each target is large. The bottom panel shows the results with the size constraint implemented. The disks are overall smaller than without the size constraint, and they have smaller ranges of sizes. Most disks fall onto the constraint, although the faintest disks exhibit a worse agreement with the trend. \citet{tripathi2017} measured the relation with disks as faint as $\log(F_\nu$/Jy) $\approx$ -1.5, and \citet{andrews2018} recently confirmed that it holds to disks that are an order of magnitude fainter. Nevertheless, our application of the size constraint to the faintest disks in our sample still represents an extrapolation of the trend. The most significant outliers above the trend include V410 X-ray 7 AB, J04334171+1750402, J04403979+2519061 AB, V819 Tau, and JH 56. The SEDs of these sources suggest that they are transition disks (or, in the case of JH 56, perhaps a cold debris disk). The need for a deficit of warm dust does not allow their SEDs to be fit with disks as small as the external constraint would require.

\subsection{Results}
\label{sec:results}

\begin{figure*}
\epsscale{1.1}
\plotone{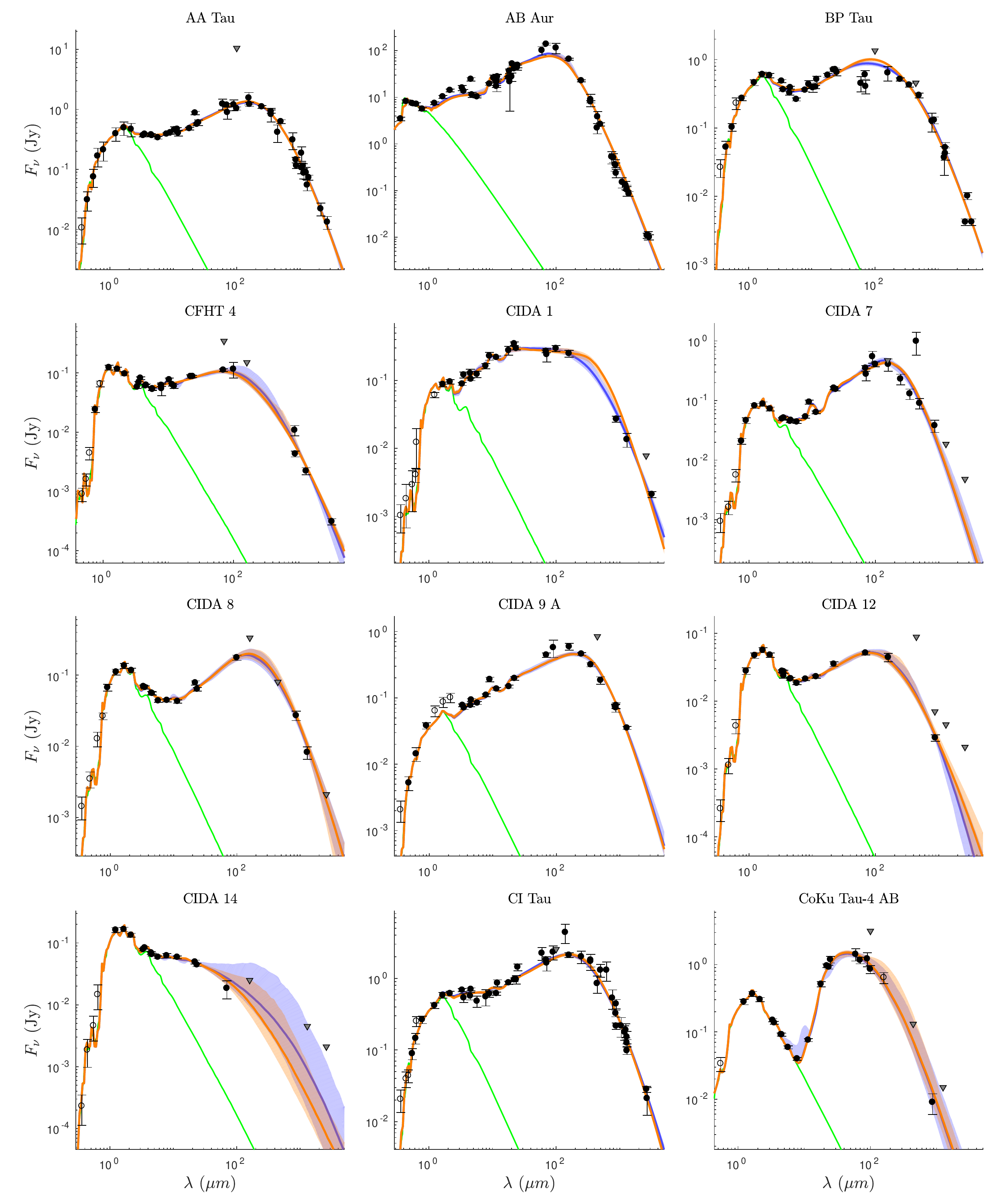}
\caption{Model fits to the measured SEDs. The data are plotted in black circles (detections) and gray triangles (3$\sigma$ upper limits). Open circles are points that were excluded from the fitting. The stellar photosphere is shown in green. The shaded regions show the range (15.9\%--84.1\%) of flux densities from the posterior sample of model fits, with the solid lines showing the median model. Models shown in blue were fit without the size constraint, while models shown in orange were fit with the size constraint. Systems marked with an asterisk were not well fit by our model and were excluded from our subsequent analysis. A subset of our targets is shown here; the complete figure set (11 images) is available in the online journal.}
\label{fig:SEDs}
\end{figure*}

The results of the MCMC fits are given in Tables \ref{table:MCMCresults-nosizeconstraint} (without the size constraint) and \ref{table:MCMCresults-withsizeconstraint} (with the size constraint). Each quoted value is the median of the posterior sample of models, with uncertainties spanning the 5.9--84.1 percentiles of the sample. We show the fits to the SEDs in Figure \ref{fig:SEDs}. Models shown in blue were fit without the size constraint, while those shown in orange were fit with the size constraint. We were not able to achieve adequate fits to \NumExcludedTargets of our targets (\ExcludedTargetsList). We exclude these targets from our demographic analysis, note them with an asterisk in Figure \ref{fig:SEDs} and Tables \ref{table:MCMCresults-nosizeconstraint} and \ref{table:MCMCresults-withsizeconstraint}, and discuss them further in the Appendix. 

Figure \ref{fig:sizeffecthistograms} shows the distributions of the median values of the eight free parameters plus $\kappa$, $T_\text{dust}$, $F_\nu / F_{\nu,\text{thin}}$, and $\log(M_\text{dust}/M_\text{dust,ana})$ from the fits both with and without the size constraint. With the size constraint, the disks are systematically smaller and warmer, and few are entirely optically thin. Without the size constraint, the $F_\nu / F_{\nu,\text{thin}}$ distribution peaks around 1 with a tail toward more optically thick disks. The other parameters show no significant difference when fitting with or without the size constraint. The dust opacities show a range of values, but the distribution is strongly peaked near the often-assumed value of 2.3 cm$^2$/g. The dust masses of most disks found from radiative transfer modeling are systematically higher than $M_\text{dust,ana}$ by a factor of $\sim$1--5. We note that other detailed SED modeling studies have also found  dust masses larger than predicted by $M_\text{dust,ana}$; e.g., \citet{mauco2018} found $M_\text{dust}/M_\text{dust,ana}$ of $\sim$3 and $\sim$6 for two disks they modeled.

\begin{figure*}
\epsscale{1.15}
\plotone{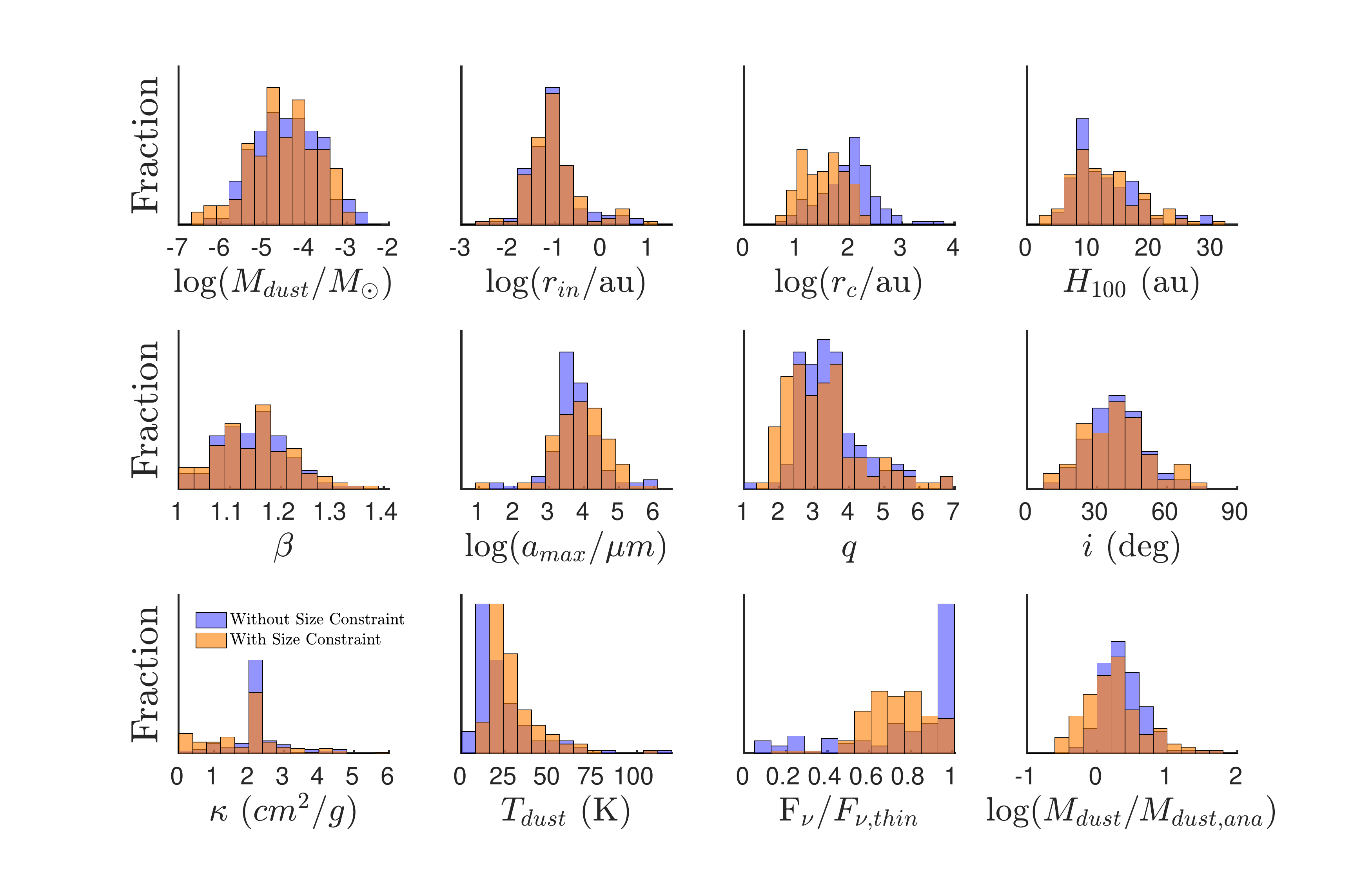}
\caption{Results of our fitting for the eight free parameters plus $\kappa$ (at 1300 $\micron$), $T_\text{dust}$, $F_\nu / F_{\nu,\text{thin}}$ (also at 1300 $\micron$), and $\log(M_\text{dust}/M_\text{dust,ana})$. The histograms show the distributions of the median values from the posterior sample of models for each target. Blue histograms are from the fits without the size constraint, and orange are from the fits with the size constraint.}
\label{fig:sizeffecthistograms}
\end{figure*}

\begin{longrotatetable}
\begin{deluxetable*}{lllllllllllll}
\tabletypesize{\scriptsize}
\tablewidth{0pt}
\tablecolumns{13}
\tablecaption{Fit Results (without Size Constraint) \label{table:MCMCresults-nosizeconstraint}}
\tablehead{\colhead{Target} & \colhead{$\log(M_\text{dust}/M_\sun)$} & \colhead{$\log(r_\text{in}/\text{au})$} & \colhead{$\log(r_c/\text{au})$} & \colhead{$H_\text{100}$ (au)} & \colhead{$\beta$} & \colhead{$\log(a_\text{max}/\micron)$} & \colhead{$q$} & \colhead{$i$ (deg)} & \colhead{$\kappa$ (cm$^2$/g)} & \colhead{$T_\text{dust}$ (K)} & \colhead{$F_\nu/F_{\nu,\text{thin}}$} & \colhead{$M_\text{dust}/M_\text{dust,ana}$}}
\startdata
AA Tau & $-4^{+0.16}_{-0.129}$ & $-0.957^{+0.1}_{-0.17}$ & $1.74^{+0.461}_{-0.206}$ & $10.3^{+1.95}_{-1.05}$ & $1.12^{+0.0184}_{-0.0157}$ & $3.26^{+0.365}_{-0.369}$ & $3.27^{+0.259}_{-0.395}$ & $29.9^{+8.86}_{-8.21}$ & $4.11^{+1.43}_{-1.07}$ & $24.9^{+6.26}_{-8.17}$ & $0.553^{+0.224}_{-0.144}$ & $1.02^{+0.447}_{-0.25}$ \\
AB Aur & $-3.84^{+0.0525}_{-0.177}$ & $-0.617^{+0.0456}_{-0.0489}$ & $2.9^{+0.116}_{-0.565}$ & $14.5^{+1.08}_{-0.493}$ & $1.14^{+0.0255}_{-0.00616}$ & $4^{+0.513}_{-2.54}$ & $4.33^{+0.107}_{-2.21}$ & $53.7^{+2.92}_{-13.3}$ & $2.18^{+0.102}_{-0.0846}$ & $30.1^{+11.8}_{-1.94}$ & $0.972^{+0.00656}_{-0.0654}$ & $2.38^{+0.225}_{-0.536}$ \\
BP Tau & $-3.97^{+0.25}_{-0.155}$ & $-1.32^{+0.104}_{-0.119}$ & $2.68^{+0.39}_{-1.52}$ & $18.3^{+3.46}_{-8.48}$ & $1.21^{+0.0309}_{-0.0932}$ & $3.33^{+0.421}_{-0.236}$ & $2.56^{+0.711}_{-0.716}$ & $28.5^{+9.7}_{-8.78}$ & $2.79^{+1.07}_{-0.733}$ & $14.2^{+28}_{-3.71}$ & $0.946^{+0.0252}_{-0.708}$ & $1.87^{+1.46}_{-0.495}$ \\
CFHT 4 & $-4.94^{+0.409}_{-0.359}$ & $-0.932^{+0.089}_{-0.228}$ & $1.78^{+1.14}_{-0.94}$ & $8.83^{+3.56}_{-3.82}$ & $1.19^{+0.0577}_{-0.0747}$ & $3.94^{+0.766}_{-2.2}$ & $3.22^{+1.8}_{-1.74}$ & $34.1^{+13.7}_{-14.2}$ & $1.82^{+0.394}_{-1.47}$ & $21.3^{+20.9}_{-12.3}$ & $0.947^{+0.0444}_{-0.308}$ & $1.9^{+3.41}_{-1.02}$ \\
CIDA 1 & $-3.82^{+0.38}_{-0.313}$ & $-0.947^{+0.0854}_{-0.1}$ & $0.801^{+2.09}_{-0.185}$ & $13.9^{+2.7}_{-1.11}$ & $1.07^{+0.0202}_{-0.0317}$ & $4.02^{+0.985}_{-1.29}$ & $4.15^{+0.978}_{-2.34}$ & $36.1^{+10.4}_{-11.5}$ & $2.07^{+0.903}_{-0.0107}$ & $45.3^{+11.1}_{-39.1}$ & $0.0987^{+0.831}_{-0.0645}$ & $4.01^{+5.03}_{-2.12}$ \\
CIDA 7 & $-4.69^{+0.482}_{-0.24}$ & $-1.04^{+0.096}_{-0.165}$ & $2.03^{+1.23}_{-0.877}$ & $11.1^{+2.19}_{-1.12}$ & $1.18^{+0.0293}_{-0.0245}$ & $4.01^{+1.63}_{-1.81}$ & $4.75^{+0.948}_{-1.24}$ & $30.2^{+21.8}_{-16.2}$ & $2.06^{+0.297}_{-0.00125}$ & $17.1^{+13.1}_{-8.06}$ & $0.945^{+0.0468}_{-0.404}$ & $1.13^{+1.61}_{-0.391}$ \\
CIDA 8 & $-4.5^{+0.421}_{-0.389}$ & $-1.24^{+0.13}_{-0.189}$ & $1.93^{+0.814}_{-0.642}$ & $7.86^{+4.4}_{-2.65}$ & $1.21^{+0.054}_{-0.0697}$ & $3.51^{+0.953}_{-1.08}$ & $3.1^{+0.633}_{-1.25}$ & $46.6^{+13.8}_{-13.5}$ & $2.22^{+2.26}_{-1.18}$ & $16.9^{+12.1}_{-7.6}$ & $0.866^{+0.107}_{-0.314}$ & $1.52^{+2.34}_{-0.88}$ \\
CIDA 9 A & $-3.99^{+0.285}_{-0.16}$ & $-1.17^{+0.061}_{-0.0844}$ & $1.77^{+0.62}_{-0.448}$ & $18.1^{+2.03}_{-1.65}$ & $1.1^{+0.0162}_{-0.0169}$ & $3.76^{+0.791}_{-0.569}$ & $3.85^{+0.44}_{-0.312}$ & $38.2^{+5.52}_{-5.66}$ & $2.77^{+0.832}_{-0.704}$ & $17.8^{+10.5}_{-7.29}$ & $0.606^{+0.256}_{-0.322}$ & $1.17^{+1.1}_{-0.379}$ \\
CIDA 12 & $-5.65^{+0.618}_{-0.412}$ & $-1.17^{+0.137}_{-0.36}$ & $1.94^{+0.999}_{-1.06}$ & $10.5^{+5.82}_{-4.83}$ & $1.25^{+0.0691}_{-0.1}$ & $2.89^{+1.26}_{-0.747}$ & $2.97^{+0.773}_{-1.36}$ & $29.7^{+21.1}_{-17.8}$ & $2.71^{+2.83}_{-1.83}$ & $16.3^{+22.6}_{-7.42}$ & $0.974^{+0.0204}_{-0.205}$ & $0.78^{+1.89}_{-0.477}$ \\
CIDA 14 & $-4.76^{+1.35}_{-1.38}$ & $-1.19^{+0.324}_{-0.351}$ & $2.36^{+0.801}_{-0.962}$ & $8.29^{+6.19}_{-5.47}$ & $1.14^{+0.0713}_{-0.08}$ & $4.08^{+1.7}_{-1.76}$ & $2.74^{+0.748}_{-0.973}$ & $50.7^{+18.2}_{-21.9}$ & $1.25^{+1.84}_{-1.21}$ & $13.6^{+16.2}_{-6.7}$ & $0.986^{+0.0114}_{-0.157}$ & $5.21^{+110}_{-4.05}$ \\
\enddata
\tablecomments{Targets marked with an asterisk did not have good fits and were excluded from our demographic analysis. (This table is available in its entirety in machine-readable form.)}
\end{deluxetable*}
\end{longrotatetable}

\begin{longrotatetable}
\begin{deluxetable*}{lllllllllllll}
\tabletypesize{\scriptsize}
\tablewidth{0pt}
\tablecolumns{13}
\tablecaption{Fit Results (with Size Constraint) \label{table:MCMCresults-withsizeconstraint}}
\tablehead{\colhead{Target} & \colhead{$\log(M_\text{dust}/M_\sun)$} & \colhead{$\log(r_\text{in}/\text{au})$} & \colhead{$\log(r_c/\text{au})$} & \colhead{$H_\text{100}$ (au)} & \colhead{$\beta$} & \colhead{$\log(a_\text{max}/\micron)$} & \colhead{$q$} & \colhead{$i$ (deg)} & \colhead{$\kappa$ (cm$^2$/g)} & \colhead{$T_\text{dust}$ (K)} & \colhead{$F_\nu/F_{\nu,\text{thin}}$} & \colhead{$M_\text{dust}/M_\text{dust,ana}$}}
\startdata
AA Tau & $-3.95^{+0.145}_{-0.12}$ & $-0.927^{+0.0776}_{-0.128}$ & $1.96^{+0.0241}_{-0.0258}$ & $10.4^{+0.883}_{-0.682}$ & $1.12^{+0.0145}_{-0.0125}$ & $3.46^{+0.328}_{-0.367}$ & $3.29^{+0.145}_{-0.188}$ & $30^{+7.97}_{-7.32}$ & $3.43^{+1.17}_{-1.05}$ & $20.4^{+0.87}_{-0.828}$ & $0.692^{+0.0107}_{-0.0131}$ & $1.18^{+0.451}_{-0.279}$ \\
AB Aur & $-3.99^{+0.0261}_{-0.0387}$ & $-0.591^{+0.0519}_{-0.0656}$ & $2.29^{+0.0407}_{-0.0598}$ & $13.4^{+0.578}_{-0.546}$ & $1.14^{+0.0164}_{-0.00909}$ & $3.19^{+1.09}_{-1.82}$ & $4.15^{+0.0995}_{-1.82}$ & $44^{+4.94}_{-3.91}$ & $2.3^{+0.16}_{-0.218}$ & $44.2^{+1.85}_{-1.4}$ & $0.888^{+0.00961}_{-0.013}$ & $1.63^{+0.17}_{-0.124}$ \\
BP Tau & $-3.97^{+0.174}_{-0.123}$ & $-1.27^{+0.0963}_{-0.101}$ & $1.83^{+0.0536}_{-0.0572}$ & $15.5^{+2.05}_{-3.24}$ & $1.18^{+0.0206}_{-0.0334}$ & $3.78^{+0.27}_{-0.238}$ & $2.57^{+0.351}_{-0.287}$ & $35.1^{+8.72}_{-11.5}$ & $1.48^{+0.578}_{-0.43}$ & $27.4^{+2.13}_{-2.54}$ & $0.744^{+0.0251}_{-0.039}$ & $1.99^{+0.982}_{-0.474}$ \\
CFHT 4 & $-5.13^{+0.385}_{-0.29}$ & $-0.927^{+0.0847}_{-0.108}$ & $1.2^{+0.103}_{-0.0996}$ & $12.3^{+4.76}_{-4.86}$ & $1.24^{+0.0433}_{-0.061}$ & $3.94^{+0.544}_{-0.866}$ & $2.57^{+0.66}_{-0.902}$ & $46.1^{+9.23}_{-12.3}$ & $1.05^{+1.07}_{-0.643}$ & $35^{+5.21}_{-7.04}$ & $0.823^{+0.0314}_{-0.0474}$ & $1.17^{+1.67}_{-0.541}$ \\
CIDA 1 & $-4.08^{+0.115}_{-0.132}$ & $-1.01^{+0.0957}_{-0.105}$ & $1.61^{+0.107}_{-0.113}$ & $13^{+3.2}_{-0.622}$ & $1.04^{+0.0237}_{-0.0161}$ & $4.22^{+1.41}_{-2.12}$ & $4.94^{+0.945}_{-1.37}$ & $26.1^{+9.66}_{-9.24}$ & $2.06^{+0.0126}_{-0.227}$ & $17.8^{+3.15}_{-2.57}$ & $0.526^{+0.109}_{-0.0944}$ & $1.81^{+0.416}_{-0.315}$ \\
CIDA 7 & $-4.84^{+0.186}_{-0.13}$ & $-1.01^{+0.0706}_{-0.108}$ & $1.4^{+0.0937}_{-0.0986}$ & $10.6^{+2.41}_{-0.784}$ & $1.19^{+0.0229}_{-0.0215}$ & $4.51^{+2.74}_{-2.52}$ & $4.8^{+0.851}_{-1.04}$ & $39^{+16.1}_{-19.3}$ & $2.06^{+0.108}_{-0.000579}$ & $25.3^{+2.68}_{-2.05}$ & $0.746^{+0.0497}_{-0.0714}$ & $0.762^{+0.117}_{-0.0985}$ \\
CIDA 8 & $-4.73^{+0.355}_{-0.397}$ & $-1.32^{+0.131}_{-0.198}$ & $1.57^{+0.12}_{-0.153}$ & $8.24^{+2.98}_{-2.34}$ & $1.21^{+0.0382}_{-0.0473}$ & $3.29^{+0.78}_{-0.871}$ & $2.96^{+0.575}_{-0.719}$ & $52^{+9.79}_{-13}$ & $2.85^{+3.08}_{-1.4}$ & $21.8^{+3.98}_{-3.13}$ & $0.713^{+0.0569}_{-0.0877}$ & $0.839^{+0.883}_{-0.456}$ \\
CIDA 9 A & $-4.06^{+0.158}_{-0.116}$ & $-1.18^{+0.0998}_{-0.0618}$ & $1.71^{+0.0452}_{-0.0542}$ & $17.5^{+1.33}_{-1.48}$ & $1.1^{+0.00767}_{-0.00669}$ & $3.89^{+0.732}_{-2.39}$ & $3.85^{+0.277}_{-0.683}$ & $39.9^{+4.37}_{-4.1}$ & $2.52^{+0.621}_{-0.446}$ & $18.3^{+0.732}_{-0.768}$ & $0.58^{+0.0358}_{-0.0657}$ & $1.05^{+0.462}_{-0.244}$ \\
CIDA 12 & $-5.35^{+1.17}_{-0.799}$ & $-1.14^{+0.133}_{-0.211}$ & $1.12^{+0.239}_{-0.26}$ & $11.4^{+4.74}_{-5.13}$ & $1.27^{+0.0473}_{-0.098}$ & $3.73^{+1.6}_{-1.37}$ & $2.18^{+1.12}_{-1.03}$ & $35.9^{+15.7}_{-16}$ & $1.65^{+2.33}_{-1.56}$ & $31.7^{+8.12}_{-8.79}$ & $0.846^{+0.0697}_{-0.109}$ & $0.935^{+12}_{-0.722}$ \\
CIDA 14 & $-4.12^{+2.37}_{-1.69}$ & $-1.19^{+0.116}_{-0.397}$ & $1.19^{+0.582}_{-0.521}$ & $8.91^{+9.48}_{-6.67}$ & $1.18^{+0.108}_{-0.112}$ & $5.7^{+2.84}_{-2.56}$ & $1.97^{+1.22}_{-1.16}$ & $42.4^{+22.2}_{-22.4}$ & $0.0151^{+2.05}_{-0.0151}$ & $47.8^{+41.1}_{-32.1}$ & $0.959^{+0.0255}_{-0.108}$ & $49.7^{+2.88e+04}_{-48.4}$ \\
\enddata
\tablecomments{Targets marked with an asterisk did not have good fits and were excluded from our demographic analysis. (This table is available in its entirety in machine-readable form.)}
\end{deluxetable*}
\end{longrotatetable}

\section{DISCUSSION}
\label{sec:discussion}

\subsection{Fidelity of Previous Assumptions}

Our modeling allows us to evaluate the assumptions that are typically employed when computing disk dust masses using Equation \ref{eq:flux-mass}. The median values of the dust opacities (at 1300 $\micron$) peak near the assumed value of 2.3 cm$^2$/g (Figure \ref{fig:sizeffecthistograms}), so this assumption is warranted. However, we find from our fitting that the uncertainty on $\kappa$ remains a significant source of uncertainty on the dust masses. Measurements at additional (sub-)millimeter wavelengths can reduce this uncertainty, and imposing independent constraints on the dust properties (e.g. from theoretical expectations) would help as well.

\begin{figure}
\epsscale{1.15}
\plotone{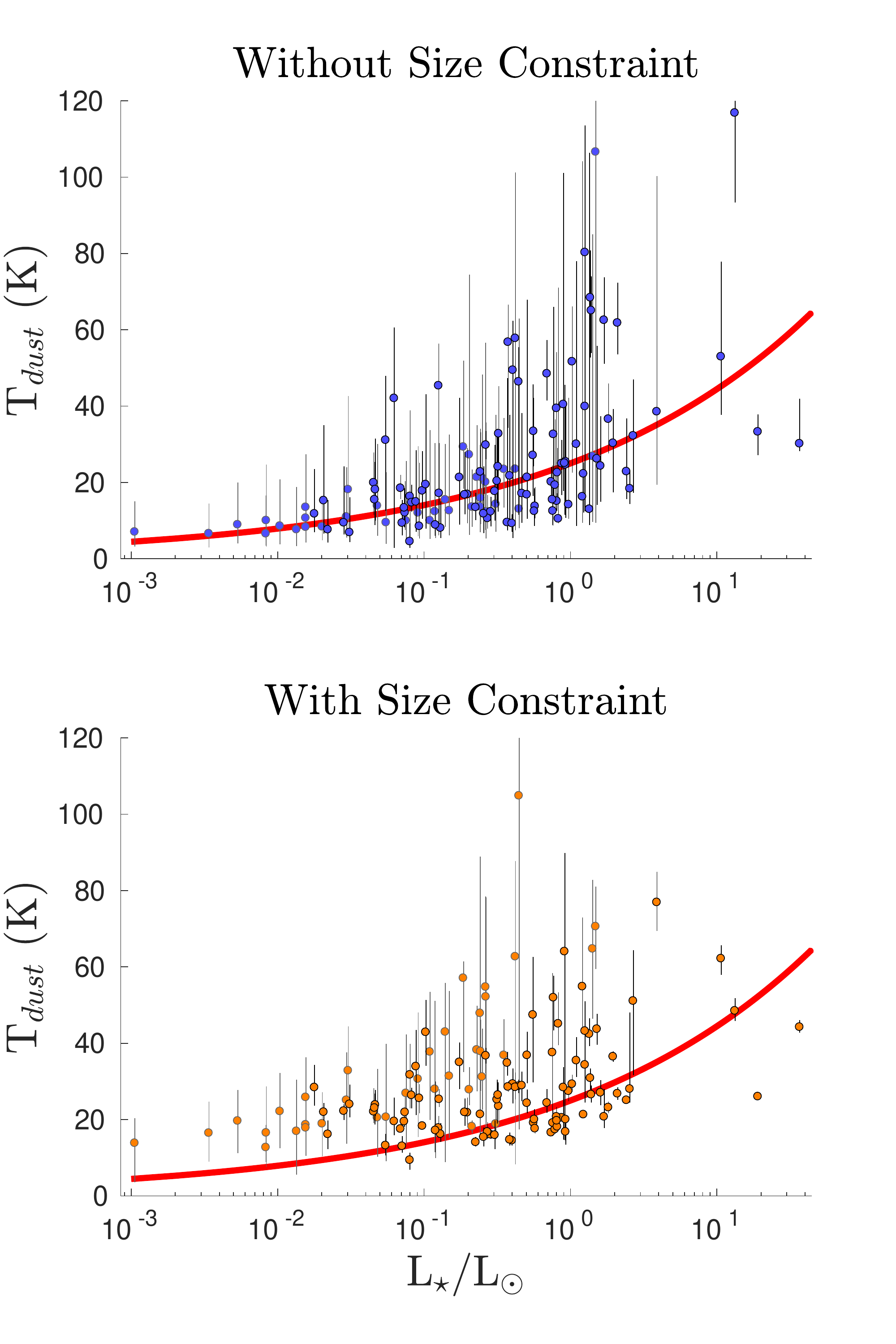}
\caption{Average dust temperatures of our model disks vs. stellar luminosity. The top panel shows the results with no size constraint, while the bottom panel shows the fits with the size constraint implemented. The points outlined in black and with black error bars are systems with detections at $\lambda >$ 500 $\micron$, while those in gray have no measurements or only upper limits at $\lambda >$ 500 $\micron$. The red line is the relation $T_\text{dust,ana} = 25 (L_\star/L_\sun)^{1/4}$ K used by \citet{andrews2013}.}
\label{fig:tdust-lstar}
\end{figure}

Radiative transfer models calculate the temperature of a disk in a realistic manner given the dust density distribution and the luminosity and spectrum of the central star, so they are well suited to test the common prescriptions for assigning a dust temperature. In Figure \ref{fig:tdust-lstar}, we plot $T_\text{dust}$ versus $L_\star$ compared with the relation for $T_\text{dust,ana}$ given in Equation \ref{eq:tdana} and used by \cite{andrews2013}. For our fits without the size constraint, the dust temperatures tend to increase with $L_\star$ and are generally consistent with the $T_\text{dust,ana}$ relation, although the uncertainties on the dust temperatures are quite large owing to the large range of disk sizes. For the fits with the size constraint, $T_\text{dust}$ also increases with $L_\star$, but the absolute temperatures overall are higher than $T_\text{dust,ana}$ because, as we showed in Figures \ref{fig:comparewithprior} and \ref{fig:sizeffecthistograms}, the size constraint forces disks to be smaller.

Radiative transfer models are also well suited to test the assumption inherent in Equation \ref{eq:flux-mass} that disk (sub-)millimeter emission is entirely optically thin. We find that when disk sizes are independently constrained, very few of them remain optically thin. Even the model fits where the disk sizes are not constrained (and the disks tend to be larger), the disks are not all optically thin. Thus, the optically thin assumption has the effect of systematically underestimating dust masses for a sample of disks.

\subsection{Planet-forming Potential}

The mass of dust in protoplanetary disks is commonly used as an indicator of their planet-forming potential. However, ensembles of disk dust masses measured using Equation \ref{eq:flux-mass} are in tension with the masses of planets seen in mature planetary systems \citep{najita2014}. Too few disks have enough dust mass to form the observed planets. A plausible solution is that the formation of planetesimals (and possibly planets) proceeds rapidly, such that a significant amount of solid mass has already been sequestered in larger bodies by the time disks reach the class II stage. The detection of annular gaps (thought to be opened by planets) in the disks of the young HL Tau system \citep{ALMA2015} and at least one embedded class I system \citep{sheehan2018} support this explanation. Our results offer another potential solution to this tension: that Equation \ref{eq:flux-mass} systematically underestimates dust masses.

\begin{figure}
\epsscale{1.2}
\plotone{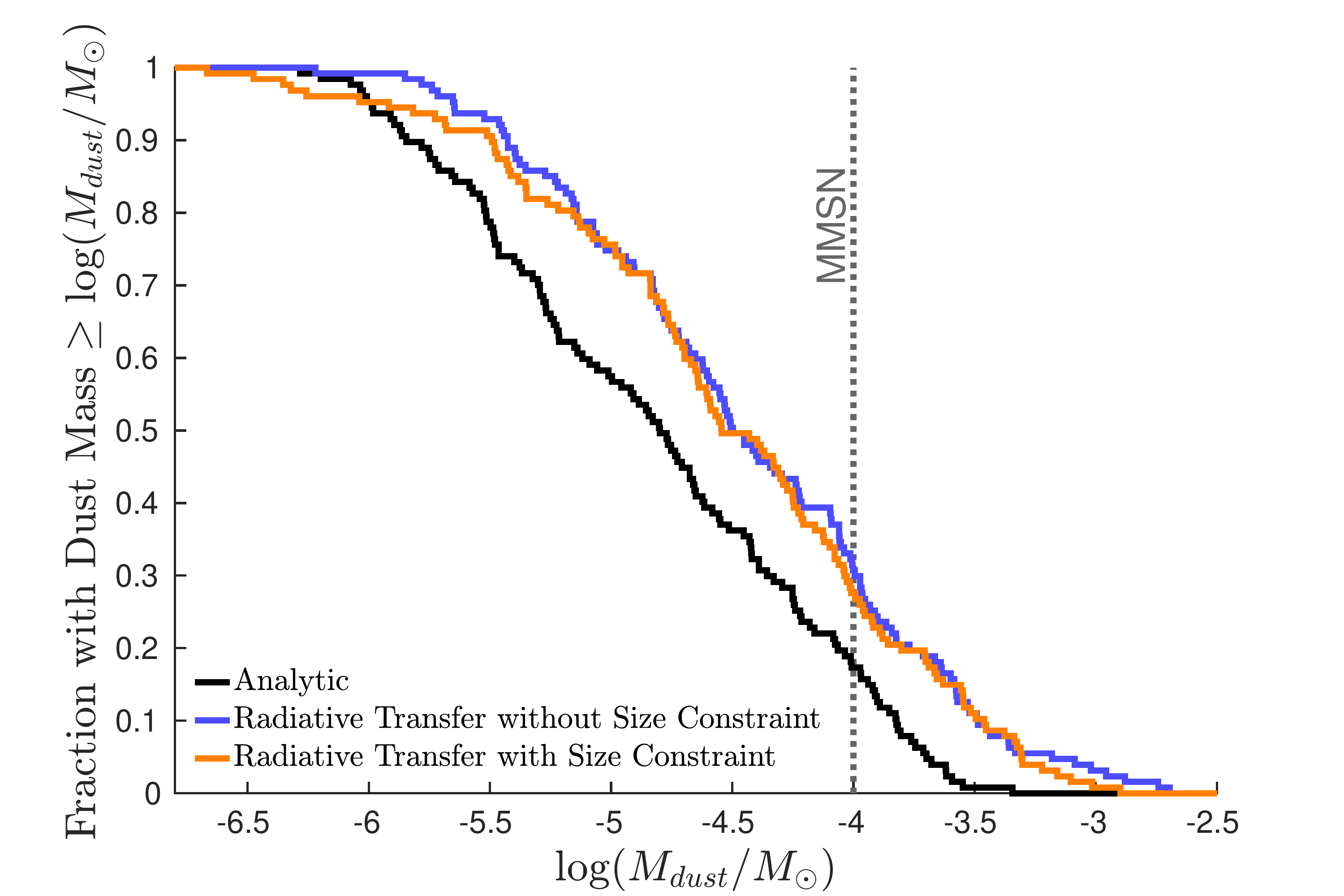}
\caption{Cumulative distribution of dust masses, i.e, the fraction of disks in our sample with dust masses at least as great as the x-axis value. The $M_\text{dust,ana}$ values are in black, and our radiative transfer models with and without the imposed size constraint are shown in orange and blue, respectively. The vertical gray dotted line indicates the minimum-mass solar nebula.}
\label{fig:MdCumDist}
\end{figure}

To quantify the planet-forming potential of our derived dust masses, we plot their cumulative distribution in Figure \ref{fig:MdCumDist}. We compare the dust masses to a benchmark value of $10^{-4} M_\sun$, a common estimate for the minimum-mass solar nebula. This value tacitly assumes a gas-to-dust ratio of 100. We find that \PercentMdAnaAboveMMSNyesconstraint\% of the systems in our Taurus sample have $M_\text{dust,ana} > 10^{-4} M_\sun$, whereas \PercentMdRadTranAboveMMSNyesconstraint\% and \PercentMdRadTranAboveMMSNnoconstraint\% have $M_\text{dust} > 10^{-4} M_\sun$ from the radiative transfer fits with and without the size constraint, respectively. The fraction of disks with dust masses of at least the minimum-mass solar nebula derived from our radiative transfer modeling are in agreement with the inferred occurrence rate for giant planets around FGK stars of $\sim\,$25\% \citep{clanton2014}. 

\subsection{Dust Mass vs. Stellar Mass}

Trends have been observed between the disk (sub-)millimeter flux density and stellar mass and between the disk dust mass and stellar mass for disks in Taurus \citep{andrews2013,ward-duong2018} and other regions \citep[e.g.][]{pascucci2016}. Here we look for these same trends in the results of our radiative transfer model fits to Taurus disk SEDs. As was done in the previous works, we used power-law fits to search for the correlations, i.e. $\log(F_\nu/\text{Jy}) = A + B \log(M_\star/M_\sun)$ and $\log(M_\text{dust}/M_\sun) = A + B \log(M_\star/M_\sun)$. The disk flux was taken at a wavelength of 1300 $\micron$. We looked for these correlations only using the model fits with the size constraint implemented.

Since we have additional information provided by our MCMC fitting routines, we are able to improve upon previous work in a couple of important ways. First, we did not use upper limits on the disk flux density or dust mass. This is because, even for targets without (sub-)millimeter detections, our models yielded dust masses and flux densities based on fits to the SED data that were available. Second, previous studies treated the uncertainties in their quantities as Gaussian errors and used a fitting algorithm appropriate for that assumption. We opted for a Monte Carlo approach to sample from the posterior set of models resulting from the MCMC fit to each target. We performed 10,000 samples. For each sample, we drew one value of our parameter of interest ($F_\nu$ or $M_\text{dust}$) and one value of $M_\star$ for each target. We used the $M_\star$ values from Table 3 of \citet{andrews2013}, which gives three different values for each target, derived using three different stellar evolution models. We performed the fit three times, once for each of these $M_\star$ determinations, as was done by \cite{andrews2013}. We sampled the $M_\star$ values by assuming that the confidence intervals described an asymmetric Gaussian distribution (the given confidence intervals were often not symmetric). We did this by first taking a 50\% chance of the value being above or below the central value, then determining the magnitude of the deviation from that central value by sampling from a Gaussian with a standard deviation being the confidence interval from the appropriate side of the distribution.

\begin{figure*}
\epsscale{1}
\plotone{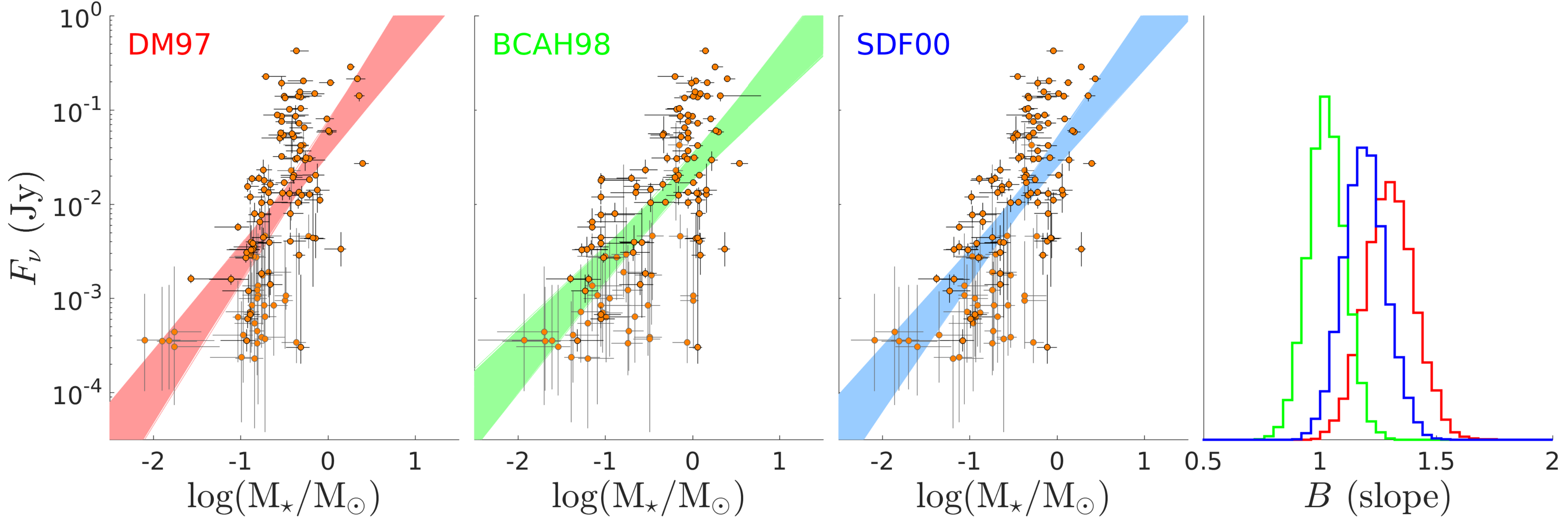}
\caption{Disk flux density at 1300 $\micron$ (from our SED fitting with the size constraint implemented) vs. stellar mass. Three different $M_\star$ determinations were used, with values taken from \citet{andrews2013}. The points outlined in black and with black error bars are systems with detections at $\lambda >$ 500 $\micron$, while those in gray have no measurements or only upper limits at $\lambda >$ 500 $\micron$. The colored regions show the distribution of power-law fits, as described in the main text. The right panel shows the histogram of power-law fit slopes.}
\label{fig:fdustmstar}
\end{figure*}

\begin{figure*}
\epsscale{1}
\plotone{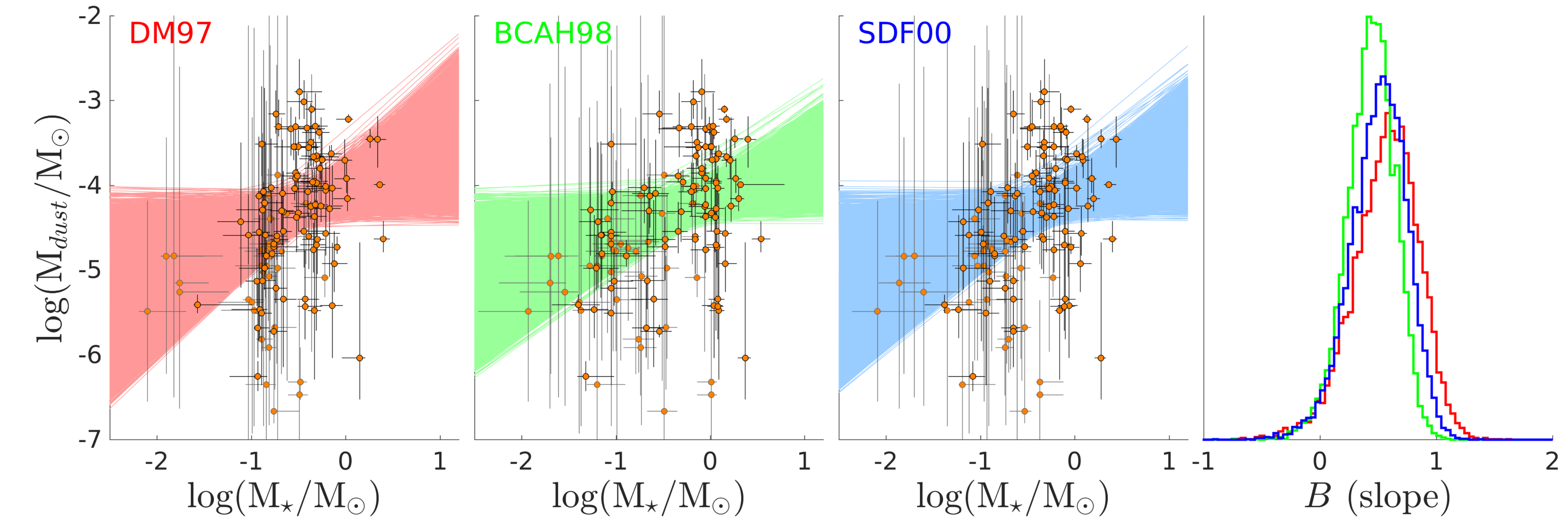}
\caption{Dust mass (from our SED fitting with the size constraint implemented) vs. stellar mass. Three different $M_\star$ determinations were used, with values taken from \citet{andrews2013}. The points outlined in black and with black error bars are systems with detections at $\lambda >$ 500 $\micron$, while those in gray have no measurements or only upper limits at $\lambda >$ 500 $\micron$. The colored regions show the distribution of power-law fits, as described in the main text. The right panel shows the histogram of power-law fit slopes.}
\label{fig:mdustmstar}
\end{figure*}

With sample values selected for each target, the best-fit values of $B$ and $A$ were computed for the power-law fit as
\begin{equation}
\label{eq:bestb}
B = \frac{\overline{\log(M_\star)\cdot\log(M_\text{dust})} - \overline{\log(M_\star)} \cdot \overline{\log(M_\text{dust})}}{\overline{\log(M_\star)^2} - \overline{\log(M_\star)}^2}
\end{equation} 
and
\begin{equation}
\label{eq:besta}
A = \overline{\log(M_\text{dust})} - B \cdot \overline{\log(M_\star)}.
\end{equation}
The same formulae were used with $M_\text{dust}$ replaced by $F_\nu$ to fit for the correlation of flux density with stellar mass. We then examined the distribution of all 10,000 $A$ and $B$ values to ascertain the statistical significance of the correlations. The results are listed in Table \ref{table:trendswithmstar}. The samples of fits are shown along with the distribution of $B$ values in Figures \ref{fig:fdustmstar} and \ref{fig:mdustmstar}. 

We find that the 1300 $\micron$ flux density does correlate with $M_\star$ to a high degree of statistical certainty, and the slope of the relation is roughly linear (or slightly steeper than linear). We are not able to recover a significant correlation between $M_\text{dust}$ and $M_\star$; a positive correlation is evident only at the 2$\sigma$ level. This is not surprising, considering previous studies mapped disk flux density to dust mass with analytic relations with the error on the flux density measurement as the primary source of uncertainty. Our fitting procedure explored many disk parameters; thus, there were many contributions to our robustly determined dust mass uncertainties.       

\begin{deluxetable*}{lccc}
\tabletypesize{\scriptsize}
\tablewidth{0pt}
\tablecolumns{4}
\tablecaption{Correlations with Stellar Mass \label{table:trendswithmstar}}
\tablehead{\colhead{} & \colhead{DM97} & \colhead{BCAH98} & \colhead{SDF00}}
\startdata
log($F_\nu$/Jy) vs. log($M_\star$/$M_\sun$) intercept (A) & -1.32 $\pm$ 0.0581 & -1.61 $\pm$ 0.0381 & -1.47 $\pm$ 0.0446 \\
log($F_\nu$/Jy) vs. log($M_\star$/$M_\sun$) slope (B) & 1.3 $\pm$ 0.104 & 1.03 $\pm$ 0.0789 & 1.19 $\pm$ 0.0905 \\
log($M_\text{dust}$/$M_\sun$) vs. log($M_\star$/$M_\sun$) intercept (A) & -4.03 $\pm$ 0.14 & -4.16 $\pm$ 0.0909 & -4.1 $\pm$ 0.108 \\
log($M_\text{dust}$/$M_\sun$) vs. log($M_\star$/$M_\sun$) slope (B) & 0.566 $\pm$ 0.284 & 0.438 $\pm$ 0.21 & 0.505 $\pm$ 0.245 \\
\enddata
\end{deluxetable*}

\section{SUMMARY}
\label{sec:summary}

\begin{enumerate}
\item We used radiative transfer models to investigate the effect of various parameters on a disk's SED, opacity, temperature, and optical depth in order to better understand the validity of the assumptions commonly used to derive dust masses from (sub-)millimeter flux density measurements.
\item The disk size and dust mass have the largest effects on the temperature and optical depth. Small disks are warmer but more optically thick---competing factors in determining the (sub-)millimeter disk flux density for a given mass.
\item The spectral indices of disks are shallower than would be expected for optically thin emission in the Rayleigh-Jeans regime. For small disks, the optical depth effects contribute most to the discrepancy, while for large disks (which are colder), the deviation from the Rayleigh-Jeans approximation contributes most to the discrepancy.
\item We fit radiative transfer models to the SEDs of \NumFinalTargets{} protoplanetary disks in Taurus. We performed the fits twice, once with unconstrained disk sizes and once with the known disk size--brightness relation imposed. The addition of the size constraint forced the disks to be smaller, warmer, and more optically thick. 
\item We found that the dust opacity values are typically peaked around the canonical value of 2.3 cm$^2$/g at 1300 $\micron$.
\item The disk temperatures show a trend with $L_\star$, as proposed by some previous studies. However, the size-constrained results have temperatures that are systematically warmer than the previously proposed relations.
\item The radiative transfer model fitting finds the dust masses to be higher than derived using millimeter photometry alone by a factor of $\sim$1--5. Using the results from radiative transfer models significantly increases the number of disks in the sample with dust masses greater than the minimum-mass solar nebula. This eases the tension between measured dust masses and the masses of planets in mature systems.
\item We recover the previously found correlation between disk millimeter flux density and stellar mass, but we only measure a positive correlation between dust mass and stellar mass at the 2$\sigma$ confidence level.
\end{enumerate}

\acknowledgments
We thank Patrick Sheehan for his help and advice on using RADMC-3D and emcee. We also thank the referee for many useful suggestions for improving the manuscript. This material is based upon
work supported by the National Aeronautics and Space
Administration under agreement No. NNX15AD94G for the
program ``Earths in Other Solar Systems". The results reported
herein benefited from collaborations and/or information
exchange within NASA's Nexus for Exoplanet System Science
(NExSS) research coordination network sponsored by NASA's
Science Mission Directorate. This work was also supported by NSF AAG grant 1311910. An allocation of computer time from the UA Research Computing High Performance Computing (HPC) at the University of Arizona is gratefully acknowledged.

\software{RADMC-3D \citep{dullemond2012}, emcee \citep{foreman-mackey2013}, DIANA Project Opacity Tool \citep{woitke2016}}

\bibliographystyle{aasjournal}

\clearpage
\appendix

\section{NOTES ON SPECIFIC TARGETS}
\label{sec:notesonsystems}

\subsection{AA Tau}
ALMA has revealed multiple radial gaps and rings in this disk \citep{loomis2017}. Furthermore, the inner part of the disk may be warped \citep{osullivan2005}. Nevertheless, our model (which assumes a smooth radial distribution) fits the SED very well with reasonable parameters, suggesting that these structures do not significantly impact the SED.  

\subsection{AB Aur}
This disk has an $r<\,$120 au cavity seen in millimeter-wavelength images, and within the cavity, there is an inner dust component and spirals of CO gas \citep{tang2012,tang2017}. Our models cannot account for this complex structure, which may explain the poor fit in the 1--10 $\micron$ region. We do achieve a good fit at longer wavelengths, suggesting that our model is accurate for the outer disk. There is also a residual envelope of material around this system, although it is unclear if this contributes significantly to the mid-IR flux \citep{lomax2016,vandermarel2016}.      

\subsection{CIDA 1}
Our fits without the size constraint favor a small disk size ($r_c\sim\,$6 au) in order to match the shape of the SED in the far-IR to submillimeter, although the confidence interval extends to much larger disks. This agrees with the SED fit by \citet{hendler2017}. Including the size constraint forces the disk size to be larger ($r_c\sim\,$40 au) with a marginally poorer fit to the SED. ALMA observations reveal this source to be a transition disk with a ring of dust peaking at $r \sim\,$20 au \citep{pinilla2018}, although perhaps with a shallow decline in the dust density toward smaller radii, which may explain the observed mid-IR excess. Future work could constrain the disk properties further by fitting the SED and ALMA visibilities simultaneously.   

\subsection{CIDA 9 A}
CIDA 9 AB is a binary system separated by 2$\farcs$3. (Sub-)millimeter imaging clearly detects emission from A and places strong upper limits on emission from B \citep{harris2012,akeson2014}; thus, we assume that only A hosts a disk. Although B is inherently only marginally less luminous than A, it suffers from $\sim\,$3 more mag of extinction \citep{andrews2013}, so we ignore the contribution from the photosphere of B on the data. We exclude the $JHK$ photometry from our fitting because it appears anomalously bright, perhaps due to the known variability of this object \citep{furlan2011}.      

\subsection{CIDA 11 AB} 
This system is a 14.1 au separation binary \citep{kraus2012}. There is no indication whether the primary or secondary star hosts the disk, or whether both stars do. Thus, we exclude this source from our sample.

\subsection{CoKu Tau-3 AB}
This system is an $\sim$ 2$\arcsec$ separation binary system \citep{kraus2012} with an ALMA-detected disk around each star \citep{akeson2014}. The separation is too close for the emission at most wavelengths to be resolved and separated into the two disks, so we exclude this source from our sample.

\subsection{CoKu Tau-4 AB}
This is an 8 au separation binary system hosting a circumbinary disk \citep{ireland2008}, so we model the stellar photosphere as a combination of the two stars. The \textit{Herschel}/PACS 160 $\micron$ photometry is potentially contaminated by large-scale nebulous emission \citep{howard2013}, so we exclude it from our fitting.

\subsection{CZ Tau AB}
This system is a $0\farcs3$ separation binary. There is no indication whether the primary or secondary star hosts the disk, or whether both stars do. Thus, we exclude this source from our sample.

\subsection{DD Tau AB}
This system is a $0\farcs3$ separation binary, and it is not clear whether one or both of the stars host a disk \citep{harris2012}. We exclude this source from our sample.

\subsection{DF Tau A}
This system is a 0$\farcs$1 separation binary system. Following the conclusion by \citet{allen2017}, we assume that only A hosts a disk. We subtracted the stellar flux contribution of B from the data before fitting the SED.

\subsection{DG Tau}
This system is very bright in the far-IR, which requires large $H_\text{100}$ values in our models, although it is also possible that these wavelengths are contaminated by nebulous envelope emission \citep{nakajima1995}.

\subsection{DH Tau A}
This is a 2$\farcs3$ separation binary system, where the secondary is a planetary-mass companion. (Sub-)millimeter emission is detected only from the primary star \citep{harris2012,wu2017,wolff2017}, so we model the system with a disk only around A and ignore any contribution to the data from the faint secondary. However, a very compact optically thick disk around the secondary could evade (sub-)millimeter detection \citep{wu2017}. Our modeling does not fit the data well in the mid-IR, where the SED exhibits a dip that may indicate a pre-transitional disk structure. Thus, we exclude this system from our demographic analysis.    

\subsection{DK Tau AB}
This system is a $3\farcs4$ separation binary system with both stars hosting disks \citep{akeson2014}. The separation is too close for the emission from the two disks to be separated at many wavelengths, so we exclude this source from our sample.

\subsection{DP Tau AB}
This is a 0$\farcs$1 binary system \citep{kraus2011}. There is no indication whether the primary or secondary star hosts the disk, or whether both stars do. Thus, we exclude this source from our sample.

\subsection{DQ Tau AB}
This system is a 0.13 au separation spectroscopic binary hosting a circumbinary disk \citep{czekala2016}. We model the stellar photosphere as a combination of the two stars.

\subsection{FM Tau}
Our models struggle somewhat to fit the 1--10 $\micron$ region of the SED. This suggests that the radial structure in the inner part of the disk may be more complicated than our model's smooth structure.

\subsection{FO Tau AB}
This is a 22 au separation binary \citep{kraus2012}. There is no indication whether the primary or secondary star hosts the disk, or whether both stars do. Thus, we exclude this source from our sample. 

\subsection{FP Tau}
The submillimeter and millimeter photometry detections of this system are clearly discrepant, and our fits effectively split the difference. Additional detections at long wavelengths would yield more decisive results for the grain size parameters. 

\subsection{FQ Tau AB}
This is a 0$\farcs$75 separation binary with (sub-)millimeter emission from both stars \citep{akeson2014}. The contribution to the SED from each disk cannot easily be separated, so we exclude this source from our sample.

\subsection{FS Tau AB}
This system is a 0$\farcs$2 separation binary \citep{harris2012}. We cannot separate contributions to the SED from potential disks around the primary, secondary, or both, so we exclude this source from our sample. For clarity, we note that there is also another star named ``FS Tau B" not in our sample and located $\sim\,$20" away that hosts an edge-on disk \citep{kirchschlager2016}.   

\subsection{FU Tau A and B}
The star FU Tau A is separated from B by 5$\farcs$7, which is too close for far-IR observations to isolate the two systems. However, far-IR observations yield only upper limits on flux density, so we apply the same upper-limit constraints to both A and B.  

\subsection{FV Tau AB}
This is a 0$\farcs$7 separation binary system with both stars likely hosting disks \citep{harris2012,akeson2014}. We cannot isolate the flux from each star at most wavelengths, so we exclude this source from our sample. 

\subsection{FV Tau-c AB}
This is a 0$\farcs$7 separation binary \citep{kraus2012}. Both stars appear to host disks, with B being a class I object \citep{mccabe2006}. We cannot isolate the flux from each star at most wavelengths, so we exclude this source from our sample. 

\subsection{FX Tau AB}
This is a 0$\farcs9$ separation binary system. Although only A is detected in the (sub-)millimeter \citep{akeson2014}, mid-IR observations suggest that B hosts a disk as well \citep{mccabe2006,skemer2011}. We cannot isolate the flux from each star at most wavelengths, so we exclude this source from our sample.

\subsection{GG Tau A}
This is a triple system where Aa-Ab are separated by 0$\farcs$24 and Ab is itself a 4 au binary \citep{difolco2014}. The Aa-Ab pair is surrounded by a disk, plus there are smaller circumstellar disks \citep{dutrey2016,yang2017}. Our model is not suitable for this complicated system, so we exclude it from our sample.   

\subsection{GG Tau B}
This is a binary system with Ba-Bb separated by 1$\farcs$48 with both stars potentially hosting disks \citep{mccabe2006}. We cannot isolate the flux from each star at most wavelengths, so we exclude this source from our sample.

\subsection{GH Tau AB}
This is a 0$\farcs$3 binary system with both stars potentially hosting disks \citep{mccabe2006}. We cannot isolate the flux from each star at most wavelengths, so we exclude this source from our sample.

\subsection{GM Aur}
The shape of the SED suggests that this system has a pre-transitional disk structure \citep{hughes2009,hornbeck2016}. There are discrepant photometry measurements in the near-to-mid-IR, likely due to variability of the inner disk \citep{espaillat2011,ingleby2015}. Our model is not able to simulate a pre-transitional disk, yet we achieved a good fit to the SED using a combination of low scale height and strong flaring. Thus, the $H_\text{100}$ and $\beta$ values we derived for this disk should be interpreted with caution.  

\subsection{GN Tau AB}
This is a $0\farcs4$ separation binary with evidence for disks around both stars \citep{skemer2011}. We cannot isolate the flux from each star at most wavelengths, so we exclude this source from our sample.   

\subsection{Haro 6-28 AB}
This is a $0\farcs7$ separation binary with both stars possibly hosting disks \citep{mccabe2006}. We cannot isolate the flux from each star at most wavelengths, so we exclude this source from our sample.   

\subsection{Haro 6-37 AB}
This is a $0\farcs3$ separation binary \citep{kraus2012}. There is no indication whether the primary or secondary star hosts the disk, or whether both stars do. Thus, we exclude this source from our sample. This system is often referred to as Aab in the literature, with the system listed here as Haro 6-37 C referred to as B.

\subsection{Haro 6-37 C}
This star is separated from Haro 6-37 AB by $2\farcs6$. This is too close to isolate the far-IR flux density from these sources, so we simply exclude the far-IR measurements from the fitting. This system is often referred to as B in the literature, with the system listed here as Haro 6-37 AB referred to as Aab.  

\subsection{HK Tau A}
This star is separated from HK Tau B by $2\farcs3$. This is too close to isolate the far-IR flux density from these sources, so we simply exclude the far-IR measurements from the fitting.

\subsection{HK Tau B}
This system hosts an edge-on disk that occults the star and inner disk \citep{stapelfeldt1998,mccabe2011}. Thus, we exclude this source from our sample.

\subsection{HN Tau A}
This is a binary system with A-B separated by 3". (Sub-)millimeter images show emission only from A \citep{harris2012,akeson2014}, so we proceed assuming that only the primary star hosts a disk.

\subsection{HP Tau}
Some far-IR measurements appear anomalously high, likely due to contamination from nebulous material around this system \citep{kirk2013}.

\subsection{HV Tau C}
This is a well-known edge-on disk \citep{stapelfeldt2003,duchene2010}, so we exclude it from our sample.

\subsection{IRAS 04173+2812}
This system may be an edge-on disk or a class I source \citep{luhman2010,furlan2011}, so we exclude it from our sample.

\subsection{IRAS 04260+2642}
This source appears to be an edge-on disk \citep{hartmann2005,furlan2011}, so we exclude it from our sample.

\subsection{IRAS 04301+2608}
This system may be an edge-on disk or a class I source \citep{furlan2011}, so we exclude it from our sample.

\subsection{IS Tau AB}
This system is a 0$\farcs$2 binary \citep{schaefer2014}. There is no indication whether the primary or secondary star hosts the disk, or whether both stars do. Thus, we exclude this source from our sample.

\subsection{ITG 33A}
This is likely an edge-on disk \citep{andrews2013}, so we exclude it from our sample. 

\subsection{IT Tau AB}
This is a 2$\farcs$4 separation binary system with both stars hosting disks \citep{harris2012,akeson2014}. We cannot isolate the flux from each star at most wavelengths, so we exclude this source from our sample.  

\subsection{J04155799+2746175}
The single (sub-)millimeter measurement is unusually bright relative to the infrared points, but our models are nevertheless able to fit the SED fairly well. Additional (sub-)millimeter measurements of this source would add clarity. 

\subsection{J04161210+2756385}
Our models struggle somewhat to fit the near-to-mid-IR region of the SED. This suggests that the radial structure in the inner part of the disk may be more complicated than our model's smooth structure.

\subsection{J04202144+2813491}
This disk is likely edge-on \citep{furlan2011,andrews2013}, so we exclude it from our sample.

\subsection{J04210795+2702204}
This source has an anomalously  bright IR excess, so we exclude it from our sample.

\subsection{J04210934+2750368}
This is a 0$\farcs8$ separation binary system \citep{cieza2012}. There is no indication whether the primary or secondary star hosts the disk, or whether both stars do. Thus, we exclude this source from our sample.

\subsection{J04213459+2701388}
We are not able to achieve a satisfactory fit to this source, so we exclude it from our demographic analysis. The \textit{Herschel}/PACS 160 $\micron$ photometry may be contaminated by nebulous emission \citep{bulger2014}, so we exclude this point from our fitting. Our model does fit the 70 $\micron$ point but does not fit the mid-IR data well, nor does it agree with the strong (sub-)millimeter upper limit. This may indicate that the 70 $\micron$ photometry is contaminated as well.    

\subsection{J04284263+2714039 AB}
This is a 0$\farcs$6 separation binary system with both stars perhaps hosting disks \citep{kraus2009}. We cannot isolate the flux from each star at most wavelengths, so we exclude this source from our sample.  

\subsection{J04290068+2755033}
The stellar temperature, luminosity, and extinction for this source are not given by \citet{andrews2013}. We use values of $T_*$ = 2700 K, $L_*$ = 0.01043 $L_\sun$, and $A_v$ = 1.71 from \citet{liu2015}.

\subsection{J04324938+2253082}
This source, also known as JH 112 B, is a 0$\farcs$5 binary with evidence for disks around both stars \citep{akeson2014}. We cannot isolate the flux from each star at most wavelengths, so we exclude this source from our sample.  

\subsection{J04330945+2246487}
The (sub-)millimeter detection of this disk appears anomalously bright, so we exclude it from our fit. Even so, we are not able to fit the SED satisfactorily with our models, so we exclude this target from our demographic analysis.

\subsection{J04381486+2611399}
This source hosts an edge-on disk \citep{luhman2007}, so we exclude it from our sample.

\subsection{J04390396+2544264}
The submillimeter and millimeter photometry detections appear somewhat discrepant. Additional detections at long wavelengths would yield more decisive results for the grain size parameters.

\subsection{J04403979+2519061 AB}
This system is an $\sim\,$7 au binary \citep{kraus2012}, so we model the system as a circumbinary disk.

\subsection{J04414489+2301513}
This source includes the Bab components of a larger quadruple system \citep{bowler2015}. The Ba-Bb separation is 0$\farcs$1. There is no indication whether the primary or secondary star hosts the disk, or whether both stars do. Thus, we exclude this source from our sample.   

\subsection{J04414825+2534304}
We note that the SED data file provided by \cite{andrews2013} is erroneously labeled as ``J04414825+2523118".

\subsection{JH 112 A}
This is a binary system with Aa-Ab separated by 1$\farcs$5 and both stars hosting disks \citep{harris2012,akeson2014}.  We cannot isolate the flux from each star at many wavelengths, so we exclude this source from our sample.

\subsection{JH 223 AB}
This is a 2" separation binary system with evidence that both stars host disks \citep{itoh2015}. We cannot isolate the flux from each star at many wavelengths, so we exclude this source from our sample.  

\subsection{JH 56}
The SED of this source resembles that of a debris disk instead of a class II protoplanetary disk.

\subsection{KPNO 3}
The submillimeter and millimeter photometry detections appear somewhat discrepant. Additional detections at long wavelengths would yield more decisive results for the grain size parameters.

\subsection{KPNO 10}
Several SED data points around $\sim$4 $\micron$ are anomalously low, and we exclude them from the fitting.

\subsection{LkHa 267}
This system either is class I or hosts an edge-on disk \citep{andrews2013}, so we exclude it from our sample.

\subsection{MHO 2 AB}
This system is a 7.3 au binary \citep{kraus2011}, so we model the system as a circumbinary disk.

\subsection{MHO 3 AB}
This system is a 4.5 au binary \citep{kraus2011}, so we model the system as a circumbinary disk.

\subsection{RW Aur AB}
This is a 1$\farcs$4 binary system where both stars host disks \citep{harris2012}. We cannot isolate the flux from each star at many wavelengths, so we exclude this source from our sample. 

\subsection{St 34 ABC}
In this system, AB forms a tight binary that is separated from C by 1$\farcs$2 \citep{kraus2011}. We do not know which of the stars host disks, so we exclude this source from our sample.

\subsection{T Tau N}
This system is separated from T Tau Sab by 0$\farcs$7, and all three stars host disks \citep{ratzka2009}. We cannot isolate the flux of T Tau N at many wavelengths, so we exclude it from our sample.

\subsection{UX Tau A}
Though part of a triple system, UX Tau A is the only one of the three stars with a disk \citep{mccabe2006}. The shape of the SED in the mid-IR suggests that this may be a pre-transtional disk \citep{espaillat2010}. Our model is not suitable for such a radial structure, and our fit is very poor. Thus, we exclude this source from our demographic analysis.  

\subsection{UY Aur AB}
This is a 0$\farcs$9 separation binary with evidence for circumstellar disks around each star \citep{akeson2014}, as well as a circumbinary disk \citep{close1998}. We cannot separate the flux from each disk at most wavelengths, so we exclude this source from our sample.  

\subsection{UZ Tau E}
This is a spectroscopic binary, which we model as a circumbinary disk around Ea+Eb. It is separated from UZ Tau W by 3$\farcs$6, which is sufficient to separate the emission from the disks at most wavelengths, except in the far-IR \citep{howard2013}. 

\subsection{UZ Tau W}
This is a 0$\farcs$4 separation binary with both stars likely hosting disks \citep{harris2012}. We cannot isolate the flux from each star at most wavelengths, so we exclude this source from our sample.

\subsection{V410 X-ray 7 AB}
This is a 4.6 au binary \citep{kraus2011}, so we model the system as a circumbinary disk.

\subsection{V807 Tau A}
This system is separated from V807 Tau Bab by 0$\farcs$3, but the Bab components do not host disks \citep{schaefer2012}. The contribution to SED data from the Bab photosphere emission has been subtracted from the measurements.

\subsection{V892 Tau AB}
This is a 7 au binary system \citep{smith2005}, so we model it as a circumbinary disk. We found a bimodal population of model fits; one population could fit the near-to-mid-IR SED, while the other could fit the far-IR-to-millimeter SED. No models could fit the whole SED well, so we exclude this source from our demographic analysis.

\subsection{V955 Tau AB}
This is a 0$\farcs$3 separation binary \citep{kraus2011} with evidence that both stars host disks \citep{mccabe2006}. We cannot isolate the flux from each star at most wavelengths, so we exclude this source from our sample. 

\subsection{VY Tau AB}
This is a 0$\farcs$7 separation binary \citep{kraus2011}. There is no indication whether the primary or secondary star hosts the disk, or whether both stars do. Thus, we exclude this source from our sample.  

\subsection{XEST 26-062}
The submillimeter and millimeter photometry detections of this system are clearly discrepant, and our fits effectively split the difference. Additional detections at long wavelengths would yield more decisive results for the grain size parameters.

\subsection{XZ Tau AB}
This is a 0$\farcs$3 separation binary \citep{kraus2011}. We do not know which of the stars host disks, so we exclude this source from our sample.

\subsection{ZZ Tau AB}
This is a 6 au binary system \citep{kraus2011}, but it does not host a circumbinary disk \citep{espaillat2012}. We do not know which of the stars host disks, so we exclude this source from our sample.

\subsection{ZZ Tau IRS}
This system likely hosts an edge-on disk \citep{furlan2011,bulger2014}, so we exclude it from our sample.

\end{document}